\documentclass[12pt]{article}
\usepackage{theorem}
\usepackage{amssymb,amsmath,mathrsfs, eucal}
\usepackage{amssymb}
\usepackage{graphicx}
\newtheorem{prop}{}[section]

{\theorembodyfont{\upshape} }
\newcommand{\boma}[1]{{\mbox{\boldmath $#1$} }}

\begin{document}
\def\X{\xi}
\def\QA{4A}
\def\QB{4B}
\def\QC{4C}
\def\QD{4D}
\def\QE{4E}
\def\QF{4F}
\def\QG{4G}
\def\QH{4H}
\def\x{{\scriptscriptstyle{X}}}
\def\y{{\scriptscriptstyle{Y}}}
\def\h{{\scriptscriptstyle{H}}}
\def\P{{\tt P}}
\def\E{{\tt E}}
\def\Y{{\tt Y}}
\def\H{{\tt H}}
\def\Ep{E_{+}}
\def\Em{E_{-}}
\def\Pp{P_{+}}
\def\Pm{P_{-}}
\def\v{v}
\def\q{q}
\def\u{u}
\def\w{w}
\def\z{z}
\def\Gi{\Lambda}
\def\Gam{\Lambda_{\dag}}
\def\kep{K}
\def\df{\boma{f}}
\def\dV{W}
\def\dQ{Q}
\def\dQQ{\QQ}
\def\sata{\ref{satell} \hspace{-0.4cm} A}
\def\satb{\hbox{\ref{satell} \hspace{-0.3cm} B}}
\def\satc{\hbox{\ref{satell} \hspace{-0.3cm} C}}
\def\satd{\hbox{\ref{satell} \hspace{-0.3cm} D}}
\def\sate{\hbox{\ref{satell} \hspace{-0.3cm} E}}
\def\satf{\hbox{\ref{satell} \hspace{-0.3cm} F}}
\def\sign{\mbox{sign}}
\def\mf{\mathfrak f}
\def\ds{\delta s}
\def\dz{\delta z}
\def\Lip{{\mathcal A}}
\def\ellu{\ell_{*}}
\def\Sgr{\Gamma_{\rho}}
\def\Sgrr{\Gamma_{\rho - \ep a}}
\def\aizn{a^i_{(0) \, n}}
\def\aiz{a^{i}_{(0)}}
\def\var{x}
\def\Zsinupa{\sin Y}
\def\Zsinupb{\sin(2 Y)}
\def\Zsinupc{\sin(3 Y)}
\def\Zsinupd{\sin(4 Y)}
\def\Zsinupe{\sin(5 Y)}
\def\Zsipama{\sin(\te - 5 Y)}
\def\Zsipamb{\sin(\te - 4 Y)}
\def\Zsipamc{\sin(\te - 3 Y)}
\def\Zsipamd{\sin(\te - 2 Y)}
\def\Zsipame{\sin(\te - Y)}
\def\Zsipanu{\sin\te}
\def\Zsipapa{\sin(\te + Y)}
\def\Sepapa{\sin(\te + Y)}
\def\Zsipapb{\sin(\te + 2 Y)}
\def\Sepapb{\sin(\te + 2 Y)}
\def\Zsipapc{\sin(\te + 3 Y)}
\def\Sepapc{\sin(\te + 3 Y)}
\def\Zsipbma{\sin(2 \te - 5 Y)}
\def\Zsipbmb{\sin(2 \te - 4 Y)}
\def\Zsipbmc{\sin(2 \te - 3 Y)}
\def\Zsipbmd{\sin(2 \te - 2 Y)}
\def\Zsipbme{\sin(2 \te - Y)}
\def\Zsipbnu{\sin (2 \te)}
\def\Zsipbpa{\sin(2 \te + Y)}
\def\Zsipbpb{\sin(2 \te + 2 Y)}
\def\Zsipbpc{\sin(2 \te + 3 Y)}
\def\Zsipcma{\sin(3 \te - 5 Y)}
\def\Zsipcmb{\sin(3 \te - 4 Y)}
\def\Zsipcmc{\sin(3 \te - 3 Y)}
\def\Zsipcmd{\sin(3 \te - 2 Y)}
\def\Zsipcme{\sin(3 \te - Y)}
\def\Zsipcnu{\sin (3 \te)}
\def\Zsipcpa{\sin(3 \te + Y)}
\def\Zsipcpb{\sin(3 \te + 2 Y)}
\def\Zsipcpc{\sin(3 \te + 3 Y)}
\def\Zsipdma{\sin(4 \te - 5 Y)}
\def\Zsipdmb{\sin(4 \te - 4 Y)}
\def\Zsipdmc{\sin(4 \te - 3 Y)}
\def\Zsipdmd{\sin(4 \te - 2 Y)}
\def\Zsipdme{\sin(4 \te - Y)}
\def\Zsipdnu{\sin (4 \te)}
\def\Zsipdpa{\sin(4 \te + Y)}
\def\Zsipdpb{\sin(4 \te + 2 Y)}
\def\Zsipdpc{\sin(4 \te + 3 Y)}
\def\Zsipema{\sin(5 \te - 5 Y)}
\def\Zsipemb{\sin(5 \te - 4 Y)}
\def\Zsipemc{\sin(5 \te - 3 Y)}
\def\Zsipemd{\sin(5 \te - 2 Y)}
\def\Zsipeme{\sin(5 \te - Y)}
\def\Zsipenu{\sin (5 \te)}
\def\Zsipepa{\sin(5 \te + Y)}
\def\Zsipepb{\sin(5 \te + 2 Y)}
\def\Zsipepc{\sin(5 \te + 3 Y)}
\def\Zsiqa{\sin(6 \te - 2 Y)}
\def\Zsiqb{\sin(6 \te)}
\def\Zsiqc{\sin(7 \te - 3 Y)}
\def\Zsiqd{\sin(7 \te - Y)}
\def\Zsiqe{\sin(8 \te - 2 Y)}
\def\Zsiqf{\sin(9 \te - 3 Y)}
\def\Zconupa{\cos Y}
\def\Zconupb{\cos(2 Y)}
\def\Zconupc{\cos(3 Y)}
\def\Zconupd{\cos(4 Y)}
\def\Zconupe{\cos(5 Y)}
\def\Zcopama{\cos(\te - 5 Y)}
\def\Zcopamb{\cos(\te - 4 Y)}
\def\Zcopamc{\cos(\te - 3 Y)}
\def\Zcopamd{\cos(\te - 2 Y)}
\def\Zcopame{\cos(\te - Y)}
\def\Zcopanu{\cos\te}
\def\Zcopapa{\cos(\te + Y)}
\def\Zcopapb{\cos(\te + 2 Y)}
\def\Zcopapc{\cos(\te + 3 Y)}
\def\Zcopbma{\cos(2 \te - 5 Y)}
\def\Zcopbmb{\cos(2 \te - 4 Y)}
\def\Zcopbmc{\cos(2 \te - 3 Y)}
\def\Zcopbmd{\cos(2 \te - 2 Y)}
\def\Zcopbme{\cos(2 \te - Y)}
\def\Zcopbnu{\cos (2 \te)}
\def\Zcopbpa{\cos(2 \te + Y)}
\def\Zcopbpb{\cos(2 \te + 2 Y)}
\def\Zcopbpc{\cos(2 \te + 3 Y)}
\def\Zcopcma{\cos(3 \te - 5 Y)}
\def\Zcopcmb{\cos(3 \te - 4 Y)}
\def\Zcopcmc{\cos(3 \te - 3 Y)}
\def\Zcopcmd{\cos(3 \te - 2 Y)}
\def\Zcopcme{\cos(3 \te - Y)}
\def\Zcopcnu{\cos (3 \te)}
\def\Zcopcpa{\cos(3 \te + Y)}
\def\Zcopcpb{\cos(3 \te + 2 Y)}
\def\Zcopcpc{\cos(3 \te + 3 Y)}
\def\Zcopdma{\cos(4 \te - 5 Y)}
\def\Zcopdmb{\cos(4 \te - 4 Y)}
\def\Zcopdmc{\cos(4 \te - 3 Y)}
\def\Zcopdmd{\cos(4 \te - 2 Y)}
\def\Zcopdme{\cos(4 \te - Y)}
\def\Zcopdnu{\cos (4 \te)}
\def\Zcopdpa{\cos(4 \te + Y)}
\def\Zcopdpb{\cos(4 \te + 2 Y)}
\def\Zcopdpc{\cos(4 \te + 3 Y)}
\def\Zcopema{\cos(5 \te - 5 Y)}
\def\Zcopemb{\cos(5 \te - 4 Y)}
\def\Zcopemc{\cos(5 \te - 3 Y)}
\def\Zcopemd{\cos(5 \te - 2 Y)}
\def\Zcopeme{\cos(5 \te - Y)}
\def\Zcopenu{\cos (5 \te)}
\def\Zcopepa{\cos(5 \te + Y)}
\def\Zcopepb{\cos(5 \te + 2 Y)}
\def\Zcopepc{\cos(5 \te + 3 Y)}
\def\e{{\scriptscriptstyle{E}}}
\def\y{{\scriptscriptstyle{Y}}}
\def\p{{\scriptscriptstyle{P}}}
\def\r{{\tt r}}
\def\ti{{\mathfrak t}}
\def\Q{Q}
\def\QQ{\mathcal Q}
\def\sc{{\scriptscriptstyle{\bullet}}}
\def\k{\boma{k}}
\def\PP{{\mathscr P}}
\def\M{A}
\def\bomu{\boma{\mu}}
\def\bonu{\boma{\nu}}
\def\bosi{\boma{\sigma}}
\def\S{S}
\def\dist{\mbox{dist}}
\def\XX{\Xi}
\def\YY{\mbox{{\rm H}}}
\def\xx{\xi}
\def\yy{\eta}
\def\yy{{\mathscr Y}}
\def\m{m}
\def\Np{\mathscr N}
\def\Lp{\mathcal L}
\def\Ti{\mathscr T}
\def\Tim{T_1}
\def\kk{K}
\def\K{{\tt K}}
\def\R{{\tt R}}
\def\RR{\mathscr{R}}
\def\Te{\Theta}
\def\ttet{\dot \Theta}
\def\A{{\tt A}}
\def\R{{\tt R}}
\def\L{{\tt L}}
\def\J{{\tt J}}
\def\dJ{{\tt S}}
\def\o{{\tt o}}
\def\oxt{\o_{\X \te}}
\def\er{{\tt r}}
\def\tee{{\tt t}}
\def\ert{\dot {\tt r}}
\def\teet{\dot {\tt t}}
\def\I{{\tt I}}
\def\Iap{{\I_{ap}}}
\def\Ia{{\I_{a}}}
\def\Tt{{\tt T}}
\def\vel{{\tt v}}
\def\mm{\sigma}
\def\aa{\alpha_{0}}
\def\aad{\alpha_{\delta}}
\def\aadp{\alpha_{\delta'}}
\def\sca{{\scriptstyle{\,\bullet\,}}}
\def\pp{{\mathfrak p}}
\def\qq{{\mathfrak q}}
\def\rr{{\mathfrak r}}
\def\ha{I^1}
\def\ka{I^2}
\def\lau{\lambda_1}
\def\lad{\lambda_2}
\def\tr{\mbox{{\rm tr}}}
\def\Gi{\Lambda}
\def\Gam{\Lambda_{\dag}}
\def\MM{\mathscr{M}}
\def\UU{\mathscr{U}}
\def\FF{\dd{\partial \overline{f} \over \partial I}}
\def\GG{\mathscr{G}}
\def\SS{\mathscr{S}}
\def\RR{\mathscr{R}}
\def\NN{\mathscr{N}}
\def\AA{\mathscr{A}}
\def\BB{\mathscr{B}}
\def\CC{\mathscr{C}}
\def\DD{\mathscr{D}}
\def\EE{\mathscr{E}}
\def\LL{\mathscr{L}}
\def\HH{\mathscr{H}}
\def\Ten{\mbox{T}}
\def\Tuu{\Ten^1_1(\reali^{d})}
\def\Tdz{\Ten^2_0(\reali^{d})}
\def\Tud{\Ten^1_2(\reali^d)}
\def\Tun{\Ten^1_n(\reali^d)}
\def\uno{1_d}
\def\ppsi{\vartheta}
\def\PPsi{\Theta}
\def\N{V}
\def\half{{1 \over 2}}
\def\II{M}
\def\JJ{N}
\def\ga{\gamma_{\scriptscriptstyle{E}}}
\def\mmu{\nu}
\def\nnu{\mu}
\def\f{g}
\def\Fs{\mathscr{G}}
\def\Kap{\mathscr{K}}
\def\STB{\scriptsize B}
\def\STBB{\scriptsize BB}
\def\STF{\scriptsize F}
\def\STFF{\scriptsize FF}
\def\TB{\scriptsize(B)}
\def\TBB{\scriptsize(BB)}
\def\TF{\scriptsize(F)}
\def\TFF{\scriptsize(FF)}
\def\XXX{\mathscr X}
\def\Piu{\mathscr P}
\def\Men{\mathscr N}
\def\ffi{\varphi}
\def\ES{{\mathcal S}}
\def\KK{{\mathscr K}}
\def\KKp{{\mathscr K}'}
\def\TT{{\mathfrak T}}
\def\kp{k'}
\def\scrscr{\scriptscriptstyle}
\def\scr{\scriptstyle}
\def\dd{\displaystyle}
\def\B{ B_{\mbox{\scriptsize{\textbf{C}}}} }
\def\Bc{ \overline{B}_{\mbox{\scriptsize{\textbf{C}}}} }
\def\ppartial{\overline{\partial}}
\def\d{d}
\def\Hinf{ H^{\infty}(\reali^d, \complessi) }
\def\Hn{ H^{n}(\reali^d, \complessi) }
\def\Hm{ H^{m}(\reali^d, \complessi) }
\def\Ha{ H^{\d}(\reali^d, \complessi) }
\def\Ld{L^{2}(\reali^d, \complessi)}
\def\Lpi{L^{p}(\reali^d, \complessi)}
\def\Lq{L^{q}(\reali^d, \complessi)}
\def\Lr{L^{r}(\reali^d, \complessi)}
\def\Knb{K^{best}_n}
\def\D{\mbox{{\tt D}}}
\def\g{ {\textbf g} }
\def\QQQ{ {\textbf Q} }
\def\AAA{ {\textbf A} }
\def\gr{\mbox{graph}~}
\def\PZ{$\mbox{P}^{0}_a$~}
\def\PZAL{$\mbox{P}^{0}_\alpha$~}
\def\PL{$\mbox{P}^{1/2}_a$~}
\def\PU{$\mbox{P}^{1}_a$~}
\def\PK{$\mbox{P}^{k}_a$~}
\def\PKU{$\mbox{P}^{k+1}_a$~}
\def\PI{$\mbox{P}^{i}_a$~}
\def\Pell{$\mbox{P}^{\ell}_a$~}
\def\PTM{$\mbox{P}^{3/2}_a$~}
\def\AZ{$\mbox{A}^{0}_r$~}
\def\AU{$\mbox{A}^{1}$~}
\def\epsilona{\epsilon^{\scriptscriptstyle{<}}}
\def\epsilonb{\epsilon^{\scriptscriptstyle{>}}}
\def\lgraffa{ \mbox{\Large $\{$ } \hskip -0.2cm}
\def\rgraffa{ \mbox{\Large $\}$ } }
\def\restriction{ \stackrel{\setminus}{~}\!\!\!\!|~}
\def\m{m}
\def\Fre{Fr\'echet~}
\def\ap{{\scriptscriptstyle{ap}}}
\def\fiap{\varphi_{\ap}}
\def\BBB{ {\textbf B} }
\def\EEE{ {\textbf E} }
\def\FFF{ {\textbf F} }
\def\TTT{ {\textbf T} }
\def\KKK{ {\textbf K} }
\def\FFi{ {\bf \Phi} }
\def\a{a}
\def\ep{\varepsilon}
\def\parn{\par\noindent}
\def\teta{M}
\def\elle{L}
\def\ro{\rho}
\def\rot{{\dot \rho}}
\def\al{\alpha}
\def\si{\sigma}
\def\vsi{\varsigma}
\def\kap{\kappa}
\def\be{\beta}
\def\de{\delta}
\def\la{{\mathfrak l}}
\def\mi{{\mathfrak v}}
\def\en{{\mathfrak n}}
\def\em{{\mathfrak m}}
\def\te{\vartheta}
\def\t{\theta}
\def\tet{\dot \vartheta}
\def\It{\dot I}
\def\Jta{J'}
\def\om{\omega}
\def\ch{\chi}
\def\complessi{{\textbf C}}
\def\reali{{\textbf R}}
\def\interi{{\textbf Z}}
\def\naturali{{\textbf N}}
\def\bT{{\textbf T}}
\def\T1{{\textbf T}^{1}}
\def\Jp{{\hat{J}}}
\def\Pip{{\hat{\Pi}}}
\def\Vp{{\hat{V}}}
\def\Fp{{\hat{F}}}
\def\Gp{{\hat{G}}}
\def\Ip{{\hat{I}}}
\def\Tp{{\hat{T}}}
\def\Mp{{\hat{M}}}
\def\La{\Lambda}
\def\Upsi{\Upsilon}
\def\Lap{{\hat{\Lambda}}}
\def\Sip{{\hat{\Sigma}}}
\def\Upsig{{\check{\Upsilon}}}
\def\Kg{{\check{K}}}
\def\ellp{{\hat{\ell}}}
\def\j{j}
\def\jp{{\hat{j}}}
\def\cir{{\scriptscriptstyle \circ}}
\def\circa{\thickapprox}
\def\vain{\rightarrow}
\def\leqs{\leqslant}
\def\geqs{\geqslant}
\def\ss{s}
\def\vains{\stackrel{\ss}{\rightarrow}}
\def\parn{\par \noindent}
\def\salto{\vskip 0.2truecm \noindent}
\def\spazio{\vskip 0.5truecm \noindent}
\def\vs1{\vskip 1cm \noindent}
\def\fine{$\Box$}
\newcommand{\rref}[1]{(\ref{#1})}
\def\beq{\begin{equation}}
\def\feq{\end{equation}}
\def\beqq{\begin{eqnarray}}
\def\feqq{\end{eqnarray}}
\def\barray{\begin{array}}
\def\farray{\end{array}}

\makeatletter \@addtoreset{equation}{section}
\renewcommand{\theequation}{\thesection.\arabic{equation}}
\makeatother
\begin{titlepage}
\begin{center}
{\huge On the averaging principle for one-frequency systems. An
application to satellite motions. }
\end{center}
\vspace{1truecm}
\begin{center}
{\large
Carlo Morosi${}^1$, Livio Pizzocchero${}^2$} \\
\vspace{0.5truecm} ${}^1$ Dipartimento di Matematica, Politecnico
di
Milano, \\ P.za L. da Vinci 32, I-20133 Milano, Italy \\
e--mail: carlo.morosi@polimi.it \\
${}^2$ Dipartimento di Matematica, Universit\`a di Milano\\
Via C. Saldini 50, I-20133 Milano, Italy\\
and Istituto Nazionale di Fisica Nucleare, Sezione di Milano, Italy \\
e--mail: livio.pizzocchero@unimi.it
\end{center}
\vspace{1truecm}
\begin{abstract}
This paper is related to our previous works \cite{uno} \cite{due} on the
error estimate of the averaging technique, for systems with one
fast angular variable. In the cited references, a general method (of
mixed analytical and numerical type) has been introduced to obtain
precise, fully quantitative estimates on the averaging error.
Here, this procedure is applied to the motion of a satellite in a
polar orbit around an oblate planet, retaining only the $J_2$ term
in the multipole expansion of the gravitational potential. To
exemplify the method, the averaging errors are estimated for the
data corresponding to two Earth satellites; for a very large
number of orbits, computation of our estimators is much less
expensive than the direct numerical solution of the equations of
motion.
\end{abstract}
\vspace{1truecm} \noindent \textbf{Keywords:} Slow and fast motions,
perturbations, averaging method.
\par \vspace{0.4truecm} \noindent \textbf{AMS 2000 Subject
classification}: 70K65, 70K70, 34C29, 70H09, 37J40, 70F15.
\end{titlepage}
\section{Introduction.}
\label{intro}
Celestial mechanics often requires approximation techniques to compute the motion of bodies over very
long times. Giving reliable error estimates on these techniques is a very practical problem,
of not simple solution. \par
In this paper we apply to a typical astronomical problem the general scheme of \cite{uno} \cite{due}
to estimate the error of the averaging principle, in the case of one fast angular variable
$\te$ (in the one-dimensional torus) and many slow variables $I = (I^i)$ (the actions).
The application we consider refers to the motion of a satellite around an oblate planet: of course,
oblateness gives a perturbation of the Keplerian, purely radial gravitational potential. There is a
classical way to express the perturbed potential as a series of spherical harmonics, where
the multipole moments of the planetary mass distribution appear as coefficients.
In the case of axial symmetry, this series only involves zonal harmonics and the $\ell$-th
term contains a multipole coefficient $J_{\ell}$
(for a reminder, see subsection \sate \, and references therein). For a slightly oblate planet, all these coefficients
are small if they are defined in a convenient, dimensionless fashion; often, $J_1 \simeq 0$ and
one retains only the $\ell=2$ term of the
previous expansion; the corresponding equations of motion for the satellite are the classical "$J_2$ problem". \par
In this paper the $J_2$ problem is considered in the restricted case of polar orbits;
this problem involves a fast
angular variable $\Te$ (the angle between the polar axis and the radius vector of the satellite)
and three slow variables $\I = (\P,\E,\Y)$ representing the parameter, the eccentricity and the
argument of the pericenter for the osculating Keplerian ellipse.
In principle, all these variables are unknown functions of the "physical" time $t$; however,
it is customary to consider as a ``time'' variable the angle itself or, rather, the ratio
$\ti$ of the angle to $2 \pi$. With this choice of the time variable, by definition the angle
evolves with the law $\Te(\ti) = [2 \pi \ti]$ (where $[~]$ indicates
equivalence mod.$2 \pi$); for obvious reasons, $\ti$ will be called
the \textsl{orbit counter}. The evolution of
$\I = (\P,\E,\Y)$ is described by a set of equations of the form
\beq {d \I^i \over d \ti} = \ep f^i(\I,[2 \pi \ti])~, \qquad \I^i(0) = I^i_0~, \label{per} \feq
where $\ep$ is a small parameter
proportional to $J_2$ (see again subsection \sate).
Independently of this specific application, a system of evolution equations like \rref{per},
with any number of unknown functions $\I = (\I^i)_{i=1,...,d}$ and a small parameter $\ep > 0$,
will be called in the sequel a \textsl{perturbed periodic system}. To any such system
one can associate an averaged system, replacing the functions $f^i$ in \rref{per}
with their averages over the angle $\Te = [2 \pi \ti]$; the solution
$\J = (\J^i)$ of the latter is in a function of $\tau := \ep \ti$. \par
The difference between the solutions $\I$
of \rref{per} and $\J$ of the averaged system, on a time scale of order $1/\ep$, can be estimated
in a fully quantitative way with the general method of \cite{uno} \cite{due}; this is what
we are going to do in the present work, for the periodic system describing the polar $J_2$ problem.
As in the cited papers, the expression "fully quantitative" means that our final estimate
will have the form
\beq | \I^i(\ti) - \J^i(\ep \ti) | \leq \ep \en^i(\ep \ti)~\qquad \mbox{for $\ti \in [0,U/\ep)$}~,
\label{formm} \feq
where $\en^i : [0,U) \vain [0,+\infty)$ are computable functions, and
$U$ is a constant, whose choice determines quantitatively
the interval of observation; we note that $U/\ep$ is the total number of
orbits. \par
The availability of an algorithm to construct the estimators $\en^i$ is
the main difference between the approach of \cite{due} and the
classical, qualitative result $\I^i(\ti) - \J^i(\ep \ti) =
O(\ep)$, see, e.g., \cite{Arn} \cite{Loc} \cite{Ver}. We are aware of the
existence of higher order versions of the averaging method,
in which the error behaves like $O(\ep^p)$ for $t
\in [0, O(1/\ep))$ ($p=2,3,...$; see again \cite{Loc}
\cite{Ver}). However, the classical theory
of higher order averaging gives little more than the qualitative
error estimate $O(\ep^p)$; perhaps an extension of our methods
could give quantitative estimators of the form $\ep^p \, \en^i_{(p)}(\ep t)$,
but this problem is left to future work.
\par Let us return to the functions $\en^i$ in Eq. \rref{formm}. Our
method to compute them for the $J_2$ problem is mainly analytical, but, in its final steps, it
requires the numerical solution of an ODE on the interval $[0,U)$;
however, this numerical computation is much faster than the direct
numerical solution of \rref{per} (a fact already appearing in
other applications, different from the $J_2$ problem, considered
in our previous papers). \par
The general treatment proposed in this
paper for the polar $J_2$ problem is subsequently exemplified,
choosing for $\ep$ and for the initial conditions $I_0 = (P_0,
E_0, Y_0)$ the values corresponding to the Earth and to the Polar
and Cos-B satellites. In this case, the estimators $\en^i$ have
been computed up to $60000$ orbits (i.e., one or two centuries),
spending few seconds of CPU time on a PC. To test their
reliability, we have computed by direct numerical integration the
differences $\I^i(\ti) - \J^i(\ep \ti)$; this is a more expensive
operation, that we have been able to perform up to $3000$ orbits
with the same PC; our  estimators are quite satisfactory on this
interval, since the functions $\ti \mapsto \ep \en^i(\ep \ti)$ are
close to the envelopes of the rapidly oscillating functions $|
\I^i(\ti) - \J^i(\ep \ti) |$. \par Of course, the treatment of
real satellites of the Earth should include, besides the $J_2$
gravitational term, other minor perturbations such as:
gravitational forces corresponding to higher order moments of the
Earth, atmospheric dragging, solar wind, tidal effects due to the
Moon gravity. All these effects could be treated, giving rise
again to a system of the form \rref{per} with more complicated
perturbation components $\ep f^i$ (even for non polar orbits: in
general, if the analysis is not restricted to the orbits in a
fixed plane, the actions are not three but five). Presumably, the
corresponding averaged system and our estimators $\en^i$ for its
error could be computed as well; this is left to future work. In
spite of the limitation to the polar $J_2$ problem, we think that
the results of this paper have some interest, because they show
that a general method for error estimates works over very long
times on a non-trivial problem. \par To conclude, we describe in
few words the organization of the paper. In Section \ref{secav},
we show how to compute the error estimators for the averaging of
any periodic system \rref{per}; this illustration specializes to
the periodic case the slightly more general setting of \cite{uno}
\cite{due} for one-frequency systems. In Section \ref{satell}, the
basic facts on the motion of a satellite in a fixed plane $\Pi$
are reviewed; the outcomes are the equations of motion for $(\I^i)
= (\P,\E,\Y)$, for any perturbation of the Keplerian potential
keeping the satellite on $\Pi$. In Section \ref{polmot}, we
specialize the previous equations to the polar $J_2$ problem; the
averaged system is solved, and we construct the algorithm to
compute the error estimators $\en^i$ $(i=\p,\e,\y)$. In Section
\ref{seces}, we perform the computations for the Polar and Cos-B
satellites of the Earth. Some details on the constructions of
Sections \ref{satell}-\ref{polmot} are presented in Appendices
\ref{appkepl}-\ref{appinv}. \salto
\section{Averaging of periodic systems.}
\label{secav}
\textbf{2A. Introducing the problem.}
We consider an open set $\Lambda$ of (the space of
the actions) $\reali^d$ and the one-dimensional torus $\bT$:
\beq \Lambda = \{ I = (I^i)_{i=1,...,d} \} \subset \reali^d~, \qquad \bT :=
\reali/ 2 \pi \interi = \{ \te \}~; \feq
the natural projection of the real axis on the torus will be written
$[~] : \reali \vain \bT$, $x \mapsto [x]$.
Now, let
\beq f = (f^i)_{i=1,...,d} \in C^\m(\Gi \times \bT, \reali^d), \quad (I,\te) \mapsto f(I,\te) \label{ef} \feq
(with $m \geqs 2$); we fix
\beq I_0 \in \Lambda~, \qquad \ep\in (0,+\infty)~, \label{io} \feq
and consider the Cauchy problem
\beq {d \I \over d \ti} = \ep f(\I,[2 \pi \ti])~, \qquad \I(0) = I_0~: \label{evol} \feq
the maximal solution (in the future) is a $C^{m+1}$ function $\I : [0,\ti_{\max}) \subset \reali
\vain \Lambda$, $\ti \mapsto \I(\ti)$
(with $\ti_{\max} \in (0,+\infty]$). Here and in the sequel, typeface symbols
are employed for \textsl{functions} of $\ti$ (or $\tau = \ep \ti)$); this allows, for
example, to distinguish the above mentioned function $\I$ from a point $I$ of $\Lambda$. \par
The averaged system associated to \rref{evol} is
\beq {d \J \over d \tau} = \overline{f}(\J)~, \qquad  \J(0) = I_0~,
\label{av} \feq
\beq \overline{f} = (\overline{f^i})_{i=1,...,d} \in C^\m(\Gi,\reali^d)~,
\qquad I \mapsto \overline{f}(I) :=
{1 \over 2 \pi} \int_{\bT} d \te~f(I,\te)~; \label{sense} \feq
the maximal solution (in the future) is a $C^{m+1}$
function $\J : [0,\tau_{\max}) \subset \reali \mapsto \Lambda~, \tau \mapsto \J(\tau)$. Throughout the paper,
we are interested in binding the difference $\I(\ti) - \J(\ep \ti)$.
\salto
\textbf{2B. Connections with the framework of \cite{uno} \cite{due}.} In these papers, we have discussed
the Cauchy problem
\beq \left\{\barray{ll} d \I/d \ti = \ep f(\I,\Te)~, &\quad \I(0) = I_0 \\
d \Te/ d \ti = \om(\I) +\ep g(\I,\Te)~, &\quad \Te(0) = \te_0 \farray
\right.~ \label{pert} \feq
with $f$ as in \rref{ef} and $g \in C^\m(\Gi \times \bT, \reali)$,~$\om \in C^\m(\Gi, \reali)$,
$\om(I) \neq 0$ for all $I \in \Gi$, $I_0 \in \Lambda, \te_0 \in \bT$, $\ep\in (0,+\infty)$,
the maximal solution being $(\I,\Te) : [0,\ti_{\max}) \vain \Lambda \times \bT$. Precise
quantitative estimates have been derived on the difference $\I(\ti) - \J(\ep \ti)$, where
$\J$ is the solution of \rref{av}. \par
Clearly, the perturbed periodic system \rref{evol} is equivalent to a particular case of \rref{pert}. In fact, let us choose
\beq \om(I) := 2 \pi~, \qquad g(I,\te) := 0 \qquad \mbox{for all $I \in \Lambda$, $\te \in \bT$}~;
\qquad \te_0 := 0~; \label{special} \feq
then, the equation for $\Te$ in \rref{pert} is fulfilled with
\beq \Te(\ti) := [ 2 \pi \ti] \feq
and  the equation for $\I$ in \rref{pert} becomes, with this position, the Cauchy problem
\rref{evol}. This remark allows to apply the general results of \cite{uno} \cite{due}
to the problem \rref{evol}; in the sequel, we report from these papers the minimal
elements enabling one to read independently the present paper, i.e.: (i) some basic
assumptions required by our approach, and the definitions of certain auxiliary functions;
(ii) the final proposition from \cite{due}, estimating $\I(\ti) - \J(\ep \ti)$ in terms of these auxiliary
functions. \salto
\textbf{2C. A first set of assumptions and auxiliary functions.} We use the tensor spaces and
the operations on tensors already employed in our
previous works; in particular, the spaces $\Tuu$, $\Tdz$ and $\Tud$ are made, respectively, by the
the $(1,1)$, $(2,0)$ and $(1,2)$ tensors over $\reali^d$; in the three cases, these are represented
as families of real numbers
$\AA = (\AA^{i}_{j})$, $\BB = (\BB^{ij})$, $\CC = (\CC^{i}_{j k})$~($i,j,k=1,...,d$). \par
From here to the end of the section the function $f$, the initial datum $I_0$ and the
parameter $\ep$ of Eqs. \rref{ef} \rref{io} are fixed. We stipulate the following:
\vskip 0.1cm\noindent
(1) $s, v \in C^\m(\Gi \times \bT,\reali^d)$,
$p, q, w \in C^{\m-1}(\Gi \times \bT, \reali^d)$,
$\u \in C^{\m-2}(\Gi \times \bT, \reali^d)$
and $\MM \in C^{\m-2}(\Gi, \Tuu)$ are the auxiliary functions uniquely defined by
the equations
\beq f = \overline{f} + 2 \pi~ {\partial s \over \partial \te}~, \quad
\overline{s} = 0~; \qquad
\label{thef} \feq
\beq s = 2 \pi {\partial \v \over \partial \te}~, \qquad
\v(I,0) = 0 \quad \mbox{for all $I \in \Gi$;} \label{st} \feq
\beq p := {\partial s \over \partial I} f~; \qquad \q := {\partial \v \over \partial I} f~;
\label{dq}  \feq
\beq p = \overline{p} + 2 \pi {\partial \w \over \partial \te}~, \qquad
\w(I,0) = 0 \quad \mbox{for all $I \in \Gi$}~; \label{qw} \feq
\beq \u := {\partial \w \over \partial I} f~; \qquad
\MM := {\partial^2 \overline{f} \over \partial I^2}\, \overline{f}  - \left(\FF\right)^2 ~. \label{em} \feq
The barred symbols $\overline{f}$, $\overline{s}$,
$\overline{p}$ denote the averages of $f,s,p$ over the angle, in the sense of \rref{sense};
$\partial/\partial I$ and $\partial^2/\partial I^2$ indicate the Jacobian and the Hessian with respect
to the variables $I = (I^i)$. Explicit formulas for $s,v$ and $w$ are
\beq s = \z - \overline{\z}~, \qquad
\z(I,\te) := {1 \over 2 \pi}~\int_{0}^{\te} d \te'~
(f(I,\te') - \overline{f}(I))~; \label{es} \feq
\beq v(I,\te) := {1 \over 2 \pi}~\int_{0}^{\te} d \te'~
s(I,\te')~; \qquad  w(I,\te) := {1 \over 2 \pi}~\int_{0}^{\te} d \te'~
(p(I,\te') - \overline{p}(I))~. \label{ev} \feq
The above definitions of $s, p, ..., \MM$ specialize the general
prescriptions of \cite{uno} \cite{due} to the case \rref{special}, of
interest in this paper. Similar comments could be added in the sequel, but they will not be repeated.
\vskip 0.1cm\noindent
(2) We introduce the open set
\beq \Gam := \{ (I,\delta I) \in \Gi \times \reali^d~|~[I, I + \delta I ]
\subset \Lambda \}~, \label{deg} \feq
where $[I, I + \delta I]$ is the closed segment in $\reali^d$ with the
indicated extremes. From now on, $\GG \in C^{\m-2}(\Gam, \Tuu)$ and $\HH
\in C^{\m-2}(\Gam,\Tud)$ are two functions such that,
for all $(I, \delta I) \in \Gam$,
\beq \overline{p}(I + \delta I) = \overline{p}(I) + \GG(I, \delta I) \, \delta I~, \label{equel0} \feq
\beq \overline{f}(I + \delta I) = \overline{f}(I) + \FF(I) \, \delta I
+ {1 \over 2} \HH(I,\delta I) \, \delta I^2,
\quad \HH^{i}_{j k}(I, \delta I) = \HH^{i}_{k j}(I, \delta I)~. \label{equel} \feq
$\GG$ and $\HH$ are not uniquely determined, if $d > 1$: possible choices, given by Taylor's formula, are
\beq \GG(I, \delta I) := \int_{0}^{1} d \var \, {\partial \overline{p}
\over \partial I}(I + \var \delta I)~,
\quad \HH(I, \delta I) := 2 \int_{0}^{1} d \var \, (1 - \var) {\partial^2
\overline{f} \over \partial I^2}(I + \var \delta I)~.
\label{tayf} \feq
(3) From now on, $[0,U)$ ($U \in (0,+\infty]$) is a fixed interval where the solution $\J$ of the averaged system \rref{av}
is assumed to exist. We denote with
$\R : [0,U) \vain \Tuu$, $\tau \mapsto \R(\tau)$ and
$\K :  [0,U) \vain \reali^d$, $\tau\mapsto \K(\tau)$
the solutions of the Cauchy problems
\beq {d \R \over d \tau} = \FF(\J) \,\R~,
\qquad \R(0) = \uno~; \label{sistr} \feq
\beq {d \K \over d \tau} = \FF(\J) \,\K + \overline{p}(\J)~,
\qquad \K(0) = 0~; \label{sistk} \feq
these are $C^\m$, and  exist on the whole interval $[0,U)$ due to the linearity
of the above differential equations. By standard arguments, one proves
that $\R(\tau)$ is an invertible matrix for all $\tau \in [0,U)$, and derives for $\K$
the explicit formula
$\K(\tau) = \R(\tau) \int_{0}^{\tau} d \tau' \, \R(\tau')^{-1} \overline{p}(\J(\tau'))$.
\salto
\textbf{2D. Some more auxiliary functions.} We maintain the assumptions and notations
of the previous items (1) (2) (3). In
\cite{due} we have introduced a second set of auxiliary functions, related to
$s, p, .... \R,\K$ and to any separating system of seminorms
on $\reali^d$. For simplicity, here we consider the seminorms giving the absolute values of
each component of vectors and tensors in $\reali^d$; what follows is an adaptation
of \cite{due} to this choice. From now on, $i, j, k$ range in $\{1,...,d\}$ and
we use Einstein's summation convention on repeated indices.
\vskip 0.1cm\noindent
(4) For $J \in \reali^d$ and $\varrho = (\varrho^i)
\in [0,+\infty]^d$, we put
\beq B(J,\varrho) := \{ I \in \reali^d~|~|I^i - J^i| < \varrho^i ~\forall~ i \}~. \feq
\vskip 0.1cm\noindent
(5) $\rho = (\rho^{i}) \in C([0,U), [0,+\infty]^d)$ is a function such that
\beq B(\J(\tau), \rho(\tau)) \subset \Lambda \qquad \mbox{for
$\tau \in [0,U)$}~. \label{recall} \feq
We put
\beq \Sgr := \{ (\tau, r)~\in [0,U) \times [0,+\infty)^d~|~
r^i < \ro^i(\tau)~\forall~ i~\}~. \label{dero} \feq
\vskip 0.1cm\noindent
(6) $a^i, b^i$ in $C^2(\Sgr, [0,+\infty))$ and
$c^i,\d^{i}_{j}, e^{i}_{j k} = e^{i}_{k j} \in C^1(\Sgr, [0,+\infty))$~
are functions such that for any $\tau \in [0,U)$,  $\delta J \in B(0, \rho(\tau))$, $\te \in \bT$,
\beq \left| \Big( s(\J(\tau) + \delta J,\te) - \R(\tau) s(I_0,\te_0) - \K(\tau) \Big)^i \right|
\leqs a^i(\tau, | \delta J|)~, \label{fa} \feq
\beq \left| \Big( \w(\J(\tau) +\delta J, \te) - \FF(\J(\tau)) \, \v(\J(\tau) +
\delta J, \te) \Big)^i \right|~
\leqs b^i(\tau, |\delta J|)~, \label{fb} \feq
\beq \Big| \Big( \u(\J(\tau)+\delta J, \te) -\FF(\J(\tau)) (\w
+\q)(\J(\tau)+\delta J,\te)   \label{fc} \feq
$$ - \MM(\J(\tau)) \v(\J(\tau)+\delta J,\te) \Big)^i \Big|
\leqs c^i(\tau,|\delta J|)~,$$
\beq | \GG^{i}_{j}(\J(\tau),\delta J)| \leqs
\d^{i}_{j}(\tau,|\delta J|)~, \label{fd} \feq
\beq | \HH^{i}_{j k}(\J(\tau),\delta J)| \leqs
e^{i}_{j k}(\tau,|\delta J|)~. \label{fe} \feq
In the above, one always intends
\beq | \delta J | := (| \delta J |^{i})_{i=1,...,d}~. \feq
The functions $c^{i}, d^{i}_{j},
e^{i}_{j k}$ are assumed to be non-decreasing with respect to the
variable $r$, i.e.,
\beq (\tau,r), (\tau, r') \in \Sgr,~~~r^{j} \leqs r'^{j}~
\forall~j ~~~\Rightarrow~~~ c^i(\tau,r) \leqs
c^i(\tau,r') \label{monot} \feq and similarly for
$d^{i}_{j}$, and $e^{i}_{j k}$. \par
\vskip 0.1cm\noindent
(7) Given $a^{i}, ...., e^{i}_{j k}$, we define the functions
\beq \alpha^{i} \in C^2(\Sgr, [0,+\infty)), \quad
\alpha^i (\tau,r) := a^{i}(\tau,r) + \ep b^{i}(\tau,r)~,\label{al} \feq
\beq \gamma^{i} \in C^1(\Sgr \times [0,+\infty)^d, [0,+\infty)), \label{ga} \feq
$$
\gamma^{i}(\tau,r,\ell) := c^{i}(\tau, r) +
\d^{i}_{j}(\tau, r) \ell^j +
{1 \over 2} e^{i}_{j k}(\tau,r) \ell^j \ell^k~.$$
\vskip 0.1cm\noindent
(8) $R^{i}_{j} \in C^1([0,U), [0,+\infty))$ and
$P^{i}_{j} \in C([0,U), [0,+\infty))$ are functions such that,
for $\tau \in [0,U)$,
\beq | \R^{i}_{j}(\tau) | \leqs R^{i}_{j}(\tau)~,~~
| (\R^{-1})^{i}_{j}(\tau) | \leqs P^{i}_{j}(\tau) ~.
\label{rp} \feq
\textbf{2E. The main result on general periodic systems.}
In \cite{due}, we have shown how to obtain estimators for $\I(\ti) - \J(\ep \ti)$
solving a system of integral equations, or of equivalent differential equations.
Here we only report the final differential reformulation, that will
be used in the present paper. Keeping in mind the previous items (1)--(8), we have the following
statement.
\begin{prop}
\textbf{Proposition.}
\label{proprinc}
(i) Assume there are $\ellu = (\ellu^i) \in [0+\infty)^d$, $(\M^i_{j}) \in [0,+\infty)^{d^2}$
and $\mm = (\mm^i) \in (0+\infty)^d$, such that
\beq \Sigma := \Pi_{i=1,...,d} [\ellu^i - \mm^i, \ellu^i + \mm^i]
\subset \Pi_{i=1,...,d} (0,\ro^i(0)/\ep)~ \label{setsi} \feq
and
\beq \Lip := \max_{i=1,...,d} \sum_{j=1}^d \M^{i}_{j} < 1/ \ep~, \label{hj} \feq
\beq \left|{\partial \alpha^i \over \partial r^j}(0, \ep \ell) \right|
\leqs \M^i_j
\qquad \mbox{for $i,j=1,...d$,~~ $\ell \in \Sigma$}, \label{hi} \feq
\beq | \alpha^i(0, \ep \ellu) - \ellu^i | + \ep \M^i_j \mm^i <
\mm^i \qquad \mbox{for $i =1,...,d$.} \label{hip} \feq
Then, there is a unique $\ell_0 = (\ell^i_0)$ such that
\beq \ell_0 \in \Sigma~,
\qquad \alpha(0, \ep \ell_0) = \ell_0~. \label{fixedp} \feq
(ii) With $\ell_0$ as above, let
$\em = (\em^i), \en = (\en^i) \in C^1([0,U),\reali^d)$ solve the Cauchy
problem
\beq {d \em^i \over d \tau} =
P^{i}_{j} \, \gamma^j(\cdot , \ep \en, \en)~, \qquad
\em^i(0) = 0~, \label{tre} \feq
\beq {d \en^i \over d \tau} = \Big(1 - \ep {\partial \alpha \over
\partial r}\, (\cdot, \ep \en)
\Big)^{-1, i}_{~~~ h}
\left( {\partial \alpha^h \over \partial \tau}\,(\cdot , \ep \en)
+ \ep R^{h}_{k} P^{k}_{j}
\, \gamma^j(\cdot , \ep \en, \en) + \ep {d R^{h}_{j} \over d
\tau} \em^j \right)~,
\label{quattro} \feq
$$ \qquad \en^i(0) = \ell^i_0  $$
for all $i$, with the domain conditions
\beq 0 < \ep \en^i < \ro^i~, \qquad
\det \Big(1 - \ep {\partial \alpha \over \partial r}\,(\cdot, \ep \en)
\Big) > 0  \label{due} \feq
(in the above, $1 - \ep {\partial \alpha/\partial r}$ stands for
the matrix $(\delta^{i}_{j} - \ep {\partial \alpha^i /\partial r^j})$
($i,j=1,...,d$), and Eq. \rref{quattro} contains the matrix elements of
its inverse.
We note that \rref{tre} implies $\em^i \geqs 0$).
Then, the solution $\I$ of the periodic system \rref{evol} exists on
$[0,U/\ep)$ and
\beq | \I^i(\ti) - \J^i(\ep \ti) | \leqs \ep \en^i(\ep t) \qquad \mbox{for $i=1,...,d$,
$\ti \in [0,U/\ep)$.} \feq
\end{prop}
\begin{prop}
\textbf{Remarks.}
\label{rema2}
\rm{
For future use, it is worthy to repeat some considerations of \cite{due}. \par \noindent
\textbf{(a)} The previous Proposition mentions $\ell_0$, the unique fixed point of the map
$\alpha(0, \ep \cdot)$ in the set $\Sigma$ of Eq. \rref{setsi}. The proof given in \cite{due}
indicates that $\alpha(0, \ep \cdot): \Sigma \vain \Sigma$ has Lipschitz constant $\ep \Lip < 1$
in the maximum component norm $\| z \| := \max_{i} | z^i |$, with $\Lip$ as in \rref{hj}. So,
from the standard theory of contractions, we have the iterative construction
\beq \ell_0 = \lim_{n \vain +\infty} l_n, \qquad l_1~ \mbox{any point of $\Sigma$}~,~
l_{n} := \alpha(0, \ep l_{n-1}) \qquad \mbox{for $n=2,3,...$}~. \label{iterates} \feq
For each $n \geqs 2$,
\beq \| \ell_0 - l_n \| \leqs (\ep \Lip)^{n-1} {\| l_2 - l_1 \| \over (1 - \ep \Lip)}~. \feq
One can use the above characterization of $\ell_0$ to compute it numerically; in this case
one finds the iterates up to a large order $n$, and then approximates $\ell_0$ with $l_n$.
\par \noindent
\textbf{(b)} In typical cases, and also in the forthcoming application to satellites, the
Cauchy problem \rref{tre} \rref{quattro} for the unknown functions $\en^i, \em^i$ is solved
numerically (by some standard package for ODEs).
}
\end{prop}
\salto
\textbf{2F. A final comment on the functions $\boma{a^i,...,e^{i}_{j k}}$.}
Let us start with a remark on $b^i, c^i, d^{i}_{j}, e^{i}_{j k}$.
These functions are always multiplied by a small factor $\ep$ in the main statements
on $| \I^i(t) - \J^i(\ep t)|$, i.e., Proposition \ref{proprinc};
for this reason, in applications it is generally sufficient to determine $b^{i},...,e^{i}_{j k}$
majorizing roughly the left-hand sides of Eqs. \rref{fb}--\rref{fe}. \par
The situation is different for the functions $a^{i}$, which are not multiplied by $\ep$; in this
case, it is important to determine them estimating accurately the left-hand side
of Eq. \rref{fa} or, at least, the part of order zero in $\delta J$.
To this purpose, we note the existence of a function $\SS \in C^{m-1}(\Gam \times \bT, \Tuu)$
such that, for all $(I, \delta I) \in \Gam$ and $\te \in \bT$,
\beq s(I + \delta I,\te) = s(I,\te) + \SS(I,\delta I,\te) \delta I \label{equels}~; \feq
a solution of this equation is given by Taylor's formula, i.e.,
\beq \SS(I, \delta I,\te) := \int_{0}^{1} d \var \, {\partial s
\over \partial I}(I + \var \delta I, \te)~. \label{tays} \feq
To go on, we write
\beq s(\J(\tau) + \delta J,\te) - \R(\tau) s(I_0,\te_0) - \K(\tau) \feq
$$ = \Big( s(\J(\tau),\te) - \R(\tau) s(I_0,\te_0) - \K(\tau) \Big) + \SS(\J(\tau), \delta J, \te) \delta J~, $$
which decomposes the function in the left-hand side of \rref{fa} into a zero order part in $\delta J$
plus a reminder controlled by $\SS$. Now, assume there are functions $\aiz \in C^2([0,U),[0,+\infty))$
and $a^{i}_{j} \in C^2(\Sgr, [0,+\infty))$ such that, for all
$\tau \in [0,U)$, $\delta J \in B(0, \rho(\tau))$ and $\te \in \bT$,
\beq \left| \Big( s(\J(\tau),\te) - \R(\tau) s(I_0,\te_0) - \K(\tau) \Big)^i \right|
\leqs \aiz(\tau)~, \label{faz} \feq
\beq | \SS^{i}_{j}(\J(\tau),\delta J, \te) | \leqs a^{i}_{j}(\tau, |\delta J|)~.
\label{faa} \feq
Then, Eq. \rref{fa} is fulfilled by the function
\beq a^i(\tau,r) := \aiz(\tau) + a^{i}_{j}(\tau,r) r^j~; \label{defamu} \feq
of course, this definition must be inserted in Eq. \rref{al} for $\gamma$. In applications
one will determine $a^{i}_{(0)}$ binding very accurately the left-hand side of Eq. \rref{faz}, and
$a^{i}_{j}$ binding more roughly the left-hand side of Eq. \rref{faa}; this is convenient, since
in Eq. \rref{defamu} for $a^i$ the terms $a^{i}_{j}$ are multiplied by $r^j$ and this will finally take
the small values $r^j = \ep \en^j(\tau)$. \par
The above strategy to determine $a^i,...,e^i_{j k}$ will be employed in the application of the
next sections, concerning satellite motions.
\section{Basic facts on satellite motions.}
\label{satell}
\textbf{\sata. A general setting.}
A satellite of mass $m$, position $P$ and velocity $\dot P$ is assumed to move around a planet of mass $M$. As a first approximation,
we assume the mass of the planet to be uniformly distributed within a sphere of
center $O$; so, the gravitational potential energy (per unit mass) of the planet has the Keplerian form
\beq V_\kep(P) = - {G M \over | P - O |} \label{vkep} \feq
where $G \simeq 6.674 \cdot 10^{-11} \, \mbox{m}^3 \mbox{Kg}^{-1}
\mbox{sec}^{-2} $ is the gravitational constant. The corresponding gravitational force
$- m \, \mbox{grad} V_\kep(P)$ will be referred to as the Keplerian force, and all the other forces
are assumed to be small perturbations of it. \par
The non-Keplerian force acting on the satellite (of gravitational, or any other
nature) will be written as
$\ep m \df(\dot P, P)$, where $\ep > 0$ is a small dimensionless parameter, arising
naturally from the analysis of this perturbation; the parameter and the satellite mass are factored out
from the perturbation, for convenience.
So, the total force acting on the satellite is
\beq \boma{F}(\dot P, P) = m \left(- \mbox{grad} V_{\kep}(P) + \ep \df(\dot P, P)\right) ~. \feq
We assume the perturbation to keep the satellite on a fixed plane $\Pi$ passing through $O$, where
we introduce the polar coordinates $\ro(P) := | P - O |$, $\te(P) :=$ angle (in $\bT$) between
a fixed unit vector $\boma{k}$ and $P - O$. The equations for the satellite motions in this plane
can be written as a first-order system for $\ro, \te$ and their derivatives $\rot, \tet$, in the
following way:
\beq  {d \rot \over d t} - \ro \tet^2 + {G M \over \ro^2} =
\ep \dQ_{\ro}(\rot, \tet, \ro, \te)~, \qquad {d \ro \over d t} = \rot~, \label{ero} \feq
\beq \rho^2 {d \tet \over d t} + 2 \ro \rot \tet = \ep \dQ_{\te}(\rot, \tet, \ro, \te)~,
\qquad {d \te \over d t} = \tet~,
\label{ete} \feq
where $\dQ_{\ro} :=  \df \boma{\cdot} (\partial P/\partial \ro)$
and $\dQ_{\te} = \df \boma{\cdot} (\partial P/\partial \te)$
are the Lagrangian components of the perturbation $\df$. In particular, in the conservative case
$\df(P) = - \mbox{grad} \dV(P)$ we have
\beq \dQ_{\ro}(\ro,\te) = - {\partial \dV \over \partial \ro}(\ro, \te)~,
\qquad \dQ_{\te}(\ro,\te) = - {\partial \dV \over \partial \te}(\ro, \te) \label{eqq}~. \feq
\salto
\textbf{\satb. Kepler elements.}
Our aim is to re-express the system \rref{ero} \rref{ete} after a change of coordinates
$(\rot, \tet, \ro,\te) \mapsto (P,E,Y,\te)$ where $P,E,Y$ are related to the unperturbed
Kepler problem (as in Eqs. \rref{ero} \rref{ete}, with $\dQ_\ro = 0$ and
$\dQ_{\te} = 0$) . For the sake of brevity, let us introduce the abbreviations
\beq \X \equiv (\rot, \tet, \ro)~, \qquad  (\X,\te) \equiv (\rot,\tet,\rho, \te)~. \feq
For given $(\X,\te)$, we define
\beq \oxt := \mbox{unperturbed Kepler orbit in $\Pi$ issuing from the initial datum $(\X, \te)$}~; \feq
as usually, we refer to this as the osculating Kepler orbit.
Let
$$ \DD := \{ (\X, \te)~|~\tet > 0~,\oxt~\mbox{is a (nondegenerate) ellipse}~\}  $$
(circular or rectilinear trajectories being excluded), and define maps
\beq  (\X, \te) \in \DD \mapsto P(\X, \te) \in (0,+\infty),
E(\X, \te) \in (0,1), Y(\X, \te) \in \bT \feq
where $P$, $E$, $Y$ are, respectively, the parameter$/R$, the eccentricity and the argument of the pericenter for the
ellipse $\oxt$; here,
\beq R := \mbox{a characteristic length of the problem} \feq
(e.g., the planetary radius)~.
In the above, the standard parameter is divided by $R$ to obtain a dimensionless quantity;
from now on, $P$ itself will be called the parameter and we will refer to $(P, E, Y,\te)$ as the Kepler
elements of $\oxt$.
As well known, $\DD$ is the set of states with $\tet > 0$ and negative unperturbed energy:
\beq \DD = \{ (\X,\te)~|~\tet > 0,~ {1 \over 2} (\rot^2 + \ro^2 \tet^2) - {G M \over \ro} < 0 \}~;
\label{defdd} \feq
furthermore,
\beq P(\X, \te) = {\ro^4 \tet^2 \over R G M}~, \label{yie1} \feq
\beq E(\X, \te) = \sqrt{1 + {2 \ro^4 \tet^2 \over G^2 M^2}
\left( {1 \over 2} \rot^2 + {1 \over 2} \ro^2 \tet^2 - {G M  \over \rho}
\right)}~, \label{yie2} \feq
\beq Y(\X, \te) := \mbox{the solution $\yy \in \bT$ of the equation} \label{yie3} \feq
$$ \mbox{$\ro =\dd{R P(\X,\te) \over 1 + E(\X,\te) \cos(\te-\yy)}$
such that $\sign \sin(\te - \yy) =  \sign \rot$}~. $$
To continue, we introduce the notation
\beq I \equiv (P,E,Y) \equiv (I^P, I^E, I^Y)~ \feq
and write $(\X,\te) \mapsto I(\X,\te)$ for the correspondence \rref{yie1}-\rref{yie3}.
The mapping
\beq \DD \vain (0,+\infty) \times (0,1) \times \bT^2~, \qquad
(\X, \te) \mapsto (I(\X,\te), \te) \label{maps} \feq
is one-to-one, with inverse
\beq (0,+\infty) \times (0,1) \times \bT^2 \vain \DD~, \label{inv0} \feq
$$ (I,\te) \mapsto (\X(I,\te), \te)~, $$
\beq \rot(I,\te) := \sqrt{{G M \over R P}} \, E \sin(\te - Y)~, \label{inv2} \feq
\beq \tet(I,\te) := \sqrt{{G M \over R^3 P^3}}~(1 + E \cos(\te - Y))^2~, \label{inv3} \feq
\beq \ro(I,\te) := {R P \over 1 + E \cos(\te - Y)} \label{inv1} \feq
(see Appendix \ref{appkepl} for more details). For future use, hereafter we give the
apocenter $\ro_{+}$, the pericenter $\ro_{-}$ and the orbital period for the unperturbed Kepler motion
on an ellipse of parameter $P$ and eccentricity $E$; these are
\beq \ro_{\pm}(P,E) = {R P \over 1 \mp E}~, \qquad
T_{orb}(P,E) = 2 \pi \sqrt{R^3 P^3 \over G M (1 - E^2)^3}~. \label{torb} \feq
Of course, the first two equations can be inverted giving
\beq P(\ro_{+}, \ro_{-}) = {2 \rho_{+} \rho_{-} \over R(\rho_{+} + \rho_{-})}~,
\qquad E(\ro_{+}, \ro_{-}) = {\rho_{+} - \rho_{-} \over \rho_{+} + \rho_{-}}~. \feq
\salto
\textbf{\satc. The equations of motions in terms of the Kepler elements.}
Using the transformations \rref{yie1}--\rref{inv1} and Eqs. \rref{ero} \rref{ete}, one
easily obtains a system of differential equations for $(I,\te)$, namely
\beq {d P \over d t} = 2 \ep \, \sqrt{P \over G M R} \, \dQQ_{\te}(I,\te)~, \label{sp} \feq
\beq {d E \over d t} = \ep \sqrt{{ R P \over G M}}~\sin(\te - Y)~{\dQQ_{\ro}(I,\te)}  \label{se} \feq
$$ + \ep \, {3 E + 4 \cos(\te - Y) + E \cos(2 \te - 2 Y) \over 2 \sqrt{G M R P}}~
{\dQQ_{\te}(I,\te)}~, $$
\beq {d Y \over d t} = - \ep \, \sqrt{{R P \over G M}} {\cos(\te - Y) \over E}~{\dQQ_{\ro}(I,\te)}  \label{sy}
\feq
$$ + \ep \, {\sin(\te - Y)(2  + E \cos(\te - Y)) \over E \sqrt{G M R P}}~{\dQQ_{\te}(I,\te)}~, $$
\beq {d \te \over d t} =  \sqrt{G M \over R^3 P^3} (1 + E \cos(\te - Y))^2~, \label{ste} \feq
\beq \dQQ_{a}(I,\te) :=
\dQ_{a}(\X(I,\te), \te)~, \quad (a \in \{\ro,\te\})~. \feq
As expected, for $\dQ_a = 0$ the above equations reproduce the fact that $P, E, Y$ are constants
of the motion for the unperturbed Kepler problem. \par
From now on, for technical reasons related to the application
of the general averaging method we will think the ``slow'' angle $Y$ as an element of $\reali$ rather than $\bT$;
so, the space of Kepler elements becomes
\beq \Lambda := \{ I \equiv (P,E,Y)~|~P \in (0,+\infty),~ E \in (0,1),~Y \in \reali \}~. \label{delambda} \feq
Taking this viewpoint amounts to identify the vector field defined by the above equations with its lift
with respect to the projection $(P,E,Y,\te) \in \Lambda \times \bT
\mapsto (P,E,[Y], \te) \in (0,1) \times (0,+\infty) \times \bT \times \bT$ (recall that
$[~] : \reali \vain \bT$ is the quotient map).
\salto
\textbf{\satd. From the physical time $\boma{t}$ to the ``orbit counter'' $\boma{\ti}$.} Let us consider the
(maximal) solution $t \in [0,t_{\max}) \mapsto (I(t),\te(t))$ of Eqs. \rref{sp}-\rref{ste}, with initial data
\beq I(0) = I_0 = (P_0,E_0,Y_0)~, \qquad \te(0) = 0~, \label{condiniz} \feq
and denote by $\ti(~): [0,t_{\max}) \mapsto \reali$, $t \mapsto \ti(t)$ the unique
$C^1$ function such that
\beq \ti(0) = 0~, \qquad \te(t) = [2 \pi \ti(t)]~. \label{ocounter} \feq
Then
\beq {d \te \over d t}(t) = 2 \pi {d \ti \over d t}(t)~, \feq
and the positivity of $d \te/d t$ ensures the function $t \mapsto \ti(t)$ to be increasing. The image
of this function is an interval $[0,\ti_{\max})$, and we denote the inverse function by
\beq t(~) : [0,\ti_{\max}) \vain [0,t_{\max})~, \qquad \ti \mapsto t(\ti)~. \feq
To go on, let us define
\beq I(\ti) := [I(t)]_{t = t(\ti)} \feq
for $\ti \in [0,\ti_{\max})$. Then $d I/d \ti = (d I/d t)/(d \ti/d t)$ $=
2 \pi (d I/d t)/(d \te/ d t)$ and Eqs. \rref{sp}-\rref{ste} with the data
\rref{condiniz} yield
\beq {d P \over d \ti} = {4 \pi \ep P^2 \over (1 + E \cos(\te - Y))^2}~
\left. {R \, \dQQ_{\te}(I,\te) \over G M}~\right|_{\te = [2 \pi \ti]}, \qquad P(0) = P_0~, \label{spte} \feq
\beq {d E \over d \ti} = {2 \pi \ep P^2 \sin(\te - Y) \over (1 + E \cos(\te - Y))^2}
~{R^2 \, \dQQ_{\ro}(I,\te) \over G M}  \label{sete} \feq
$$ +  \pi \ep \left. P~{3 E + 4 \cos(\te - Y) + E \cos(2 \te - 2 Y) \over (1 + E \cos(\te - Y))^2}~
{R \, \dQQ_{\te}(I,\te) \over G M}~\right|_{\te = [2 \pi \ti]}, \qquad E(0) = E_0~, $$
\beq {d Y \over d \ti} = - 2 \pi \ep {P^2 \cos(\te - Y) \over E (1 + E \cos(\te - Y))^2}~
{R^2 \, \dQQ_{\ro}(I,\te) \over G M}   \label{syte}
\feq
$$ + 2 \pi \ep \left. {P \sin(\te - Y)(2  + E \cos(\te - Y) \over E (1 + E \cos(\te - Y))^2}~
{R \, \dQQ_{\te}(I,\te) \over G M}~\right|_{\te = [2 \pi \ti]}, \qquad Y(0) = Y_0~;  $$
we note that $\ti$, $R^2 \, \dQQ_{\ro}(I,\te)/(G M)$ and $R \, \dQQ_{\te}(I,\te)/(G M)$
are dimensionless quantities. Eqs. \rref{spte}--\rref{syte}
are a system of the general form \rref{evol} for the function $\ti \mapsto I(\ti)$.
Eq. \rref{ocounter} indicates that $\ti(t)$ is the number of orbits performed
by the satellite from time $0$ to time $t$; this explains the name of \textsl{orbit counter} employed
for the $\ti$ variable.
\salto
\textbf{\sate. Polar motions of a satellite around an oblate planet.}
From now on, the perturbation is due to a deviation of the planetary mass from the
uniform distribution inside a sphere. The mass distribution is assumed to be symmetric with respect to the polar axis;
$O$ is the midpoint between the poles, and we introduce a system of Cartesian coordinates
$x,y,z$ with origin $O$, with the $z$ axis (indicated by a unit vector $\k$) as polar axis
({\footnote{Of course,
we are assuming $m/M$ to be so small that $O$ can be regarded at rest.}}).
The total gravitational potential $P \mapsto V(P)$ produced by the planet
admits a well known expansion in zonal harmonics, namely,
\beq V(P) = - {G M \over |P - O|} \left[ 1 - \sum_{\ell=1}^{+\infty} {R^\ell \over | P - O |^\ell}
J_{\ell} P_{\ell}\Big({z(P) \over |P- O|}\Big) \right] \label{geopot} \feq
where $P_{\ell}$ are the Legendre polynomials, and
\beq J_{\ell} := - {1 \over M R^\ell} \int P_{\ell}\Big({z(Q) \over |Q-O|}\Big) \, | Q - O |^\ell \, d M(Q)~. \feq
In the above: $M$ is total mass of the planet; $R$ is a typical length expressing the size of
the planet, usually chosen as the equatorial radius; $P_{\ell}$ are the Legendre polynomials;
$d M$ is the measure
describing the mass distribution of the planet.
We note that $J_{1}, J_{2}, J_{3},...$ are
dimensionless due to the presence of the parameters $M$ and $R$ in their definitions; we will refer to them
as the (dimensionless) dipole, quadrupole, octupole,... coefficients.
For a derivation of \rref{geopot} one can refer, e.g., to \cite{Kel}.
\par
As for the applications of this
expansion to (polar or generic) satellite motions, there is an enormous literature;
we only cite the classical references \cite{Ver} and \cite{Kyn}.
To continue, let us recall that
\beq P_{1}(\zeta) := \zeta~, \qquad P_2(\zeta) := {1 \over 2} (3 \zeta^2 - 1)~; \feq
in general $P_{\ell}(-\zeta) = (-1)^\ell P_{\ell}(\zeta)$, which implies $J_{\ell} = 0$ if $\ell$ is
odd and the mass distribution is reflection-invariant with respect to the equatorial plane.
Often, $J_1 \simeq 0$ even without an exact equatorial symmetry; in this case,
the first non-negligible contribution to the expansion \rref{geopot} is the $\ell=2$ term. From now on,
we assume $J_1=0$ and neglect the terms of \rref{geopot} with $\ell \geqs 3$; explicitating $P_2$, we
finally obtain
\beq V(P) = V_{\kep}(P) + \ep W(P)~, \label{vp} \feq
where $V_\kep(P) = - G M/|P - O|$ is the Keplerian potential and
\beq W(P) :=  - {G M R^2 \over | P - O |^3}~\left(1 - 3 {z^2(P) \over |P - O|^2}\right)~, \label{wp} \feq
\beq \ep := {J_2 \over 2} =
{1 \over 4 M R^2} \int \left(|Q - O|^2 - 3 z^2(Q)\right) d M(Q)~. \feq
In the sequel we always regard Eq.
\rref{vp} as giving the exact potential in the region where the satellite is
moving, and assume $\ep > 0$
({\footnote{As an example, suppose the planet is an ellipsoid $x^2/R^2 + y^2/R^2 + z^2/(\alpha R)^2
\leq 1$ and the mass distribution is uniform, i.e., $d M = (M/V) d V$ with $V = (4/3) \pi \alpha R^3$
the total volume. Then, passing to cylindrical coordinates $r, \varphi, z$ we obtain
$$ \ep = {M/V \over 4 M R^2} \left(\int_{-\alpha R}^{\alpha R} \! \! \! d z \int_{0}^{\sqrt{R^2 - z^2/\alpha^2}}
\hspace{-0.5cm} d r \, r
\int_{0}^{2 \pi} \! \! \! d \varphi~(r^2 - 2 z^2) \right)= {1 - \alpha^2 \over 10}~, $$
which implies $\ep > 0$ if $\alpha < 1$. As a supplementary remark, we mention that the analogous
of $W$ employed in \cite{Ver} has a wrong sign.}}). The values of $\ep$ when the planet is
the Earth will be reported later on. \par
By a \textsl{polar motion} of the satellite, we
mean one with initial position and velocity in a plane $\Pi$
containing the polar axis.
Such a motion stays in $\Pi$ for all times, since the gravitational force
at any point of $\Pi$ is parallel to this plane.
To analyze the motion we use on $\Pi$ a system of polar coordinates
$(\rho,\te)$, with $\te$ the angle from $\k$.
In these coordinates,
\beq W(\rho,\te) = - {G M R^2 \over \ro^3}~\left(1 - 3 \, \cos^2 \te \right)~, \feq
and the motion is described by Eqs. \rref{ero}--\rref{eqq}.
A polar motion in the domain $\DD$ of Eq. \rref{defdd} can be described via
the Kepler elements $(I,\te) = (P,E,Y,\te)$, and we can write down Eqs. \rref{sp}--\rref{ste}
or \rref{spte}--\rref{syte} for the present choice of the perturbation.
The explicit form of Eqs. \rref{spte}--\rref{syte} is
\beq {d P \over d \ti} = \ep f^\p(I,[2 \pi \ti]), \quad
{d E \over d \ti} = \ep f^\e(I,[2 \pi \ti]), \quad
{d Y \over d \ti} = \ep f^\y(I,[2 \pi \ti]), \label{eqpey} \feq
$$ P(0) = P_0~, \qquad E(0) = E_0~, \qquad Y(0) = Y_0~, $$
depending on the function
\beq f = (f^i)_{i=\p,\e,\y} : \Lambda \times \bT \vain \reali^3~,
\qquad (I,\te) \mapsto f(I,\te)~, \label{fp} \feq
$$ f^\p(I,\te) := {6 \pi \over P}~\Big(  E \sin(\te + Y) + 2 \sin(2 \te) + E \sin(3 \te - Y) \Big)~,  $$
$$ f^\e(I,\te) :=
{3 \pi \over 8 P^2}~\Big(E^2 \sin(\te - 3 Y) + (8 + 2 E^2) \sin(\te - Y)
+ (4 + 11 E^2) \sin(\te + Y)  $$
$$ + 8 E \sin(2 \te - 2 Y) + 40 E \sin(2 \te)  +
2 E^2 \sin(3 \te - 3 Y)   $$
$$  + (28 + 17 E^2) \sin(3 \te - Y) + 24 E \sin(4 \te - 2 Y) + 5 E^2 \sin(5 \te - 3 Y)~\Big)~, $$
$$ f^\y(I,\te) := - {3 \pi \over P^2} -
{3 \pi \over 8 E P^2}~\Big(E^2 \cos(\te - 3 Y)
+ (8 + 6 E^2) \cos(\te - Y)  $$
$$ - (4 - 7 E^2) \cos(\te + Y) + 8 E \cos(2 \te - 2 Y)
+ 24 E \cos(2 \te) + 2 E^2 \cos(3 \te - 3 Y)  $$
$$ + (28 + 11 E^2) \cos(3 \te - Y)  + 24 E \cos(4 \te - 2 Y)
+ 5 E^2 \cos(5 \te - 3 Y) \Big)~. $$
This is the system to which we will apply the general scheme of Section \ref{secav}.
\section{Polar motions around an oblate planet: the averaging method and its error.}
\label{polmot}
\salto
We take as a starting point the formulation of the problem in terms
of the variables $I = (P,E,Y) \in \Lambda := (0,+\infty) \times (0,1) \times \reali$,
based on the orbit counter $\ti$ as independent variable. So, the evolution
equations have the form \rref{evol} with $f = (f^\p, f^\e, f^\y)$ as in Eqs.
\rref{fp}. From now on, for consistency with the general notation of Sections
\ref{intro} and \ref{secav}, we write
\beq \I = (\P,\E,\Y) \feq
for the (maximal) solution of these equations for given initial data (we repeat that
typeface symbols denote functions of $\ti$ or $\tau = \ep \ti$; thus $\I : \ti \mapsto \I(\ti)$ is a function,
to be distinguished from a point $I = (P,E,Y)$ of the
space $\Lambda$). Throughout the section, indices $i, j, h, k$ range in $\{\p,\e,\y\}$. \parn
In the sequel, we compute the averaged equations of motion and their
solutions; then we will evaluate the error of averaging with the general method
of Section \ref{secav}. \salto
\textbf{\QA. The averaged system.} The averages over $\te \in \bT$ of the functions
\rref{fp} are $\overline{f^i}$, where, for all $I = (P,E,Y)$,
\beq \overline{f}^{\, \p}(I) = 0~, \quad \overline{f}^{\, \e}(I) = 0~, \quad
\overline{f}^{\, \y}(I) = - {3 \pi \over P^2}~. \label{mf} \feq
So, the averaged system for $\J = (\J^\p, \J^\e, \J^\y)$ is
\beq {d \J^\p \over d \tau} = 0,~{d \J^\e \over d \tau} = 0,~
{d \J^\y \over d \tau} = - {3 \pi \over (\J^\p)^2}, \quad (\J^\p, \J^\e, \J^\y)(0) = (P_0, E_0, Y_0) \in \Lambda~, \feq
and has solution
\beq \J^\p(\tau) = P_0~, \qquad \J^\e(\tau)= E_0~, \qquad \J^\y(\tau) = Y_0 - {3 \pi \over {P_0}^2} \, \tau
\quad \mbox{for $\tau \in [0,+\infty)$}\label{solmed} \feq
(which stays in $\Lambda$ for all $\tau$). As we see, in the approximation given
by averaging, the parameter and eccentricity are
constant, while the argument of the pericenter varies linearly with the rescaled time $\tau =
\ep \ti$; this is a classical result. \par
\salto
\textbf{\QB. The auxiliary functions $\boma{s,v,..., \FF, \MM, \SS,}$ $\boma{\GG, \HH}$.} These are
required by our general method to evaluate the error of the averaging method; their definitions
are found in Eqs. \rref{thef}--\rref{em}.
We have computed all these functions by means of MATHEMATICA. \par
In the case of $s$, the components are
\beqq & s^\p&\!\!\!\!\!(I,\te) = - {1 \over P}~\Big[ 3 E \cos (\te+ Y) +
3 \cos (2\te) + E \cos (3\te- Y) \Big]~,
 \label{sx} \\
& s^\e&\!\!\!\!\!(I,\te) = - {1 \over 16 P^2} \Big[3 E^2 \cos(\te - 3 Y) +
(24 + 6 E^2) \cos(\te - Y)  + (12 + 33 E^2) \Zcopapa  \nonumber \\
& + & \! \! \! 12 E \Zcopbmd + 60 E \Zcopbnu +   2 E^2 \Zcopcmc  + (28 + 17 E^2) \Zcopcme  \nonumber \\
& + & \! \! \!  18 E \Zcopdmd + 3 E^2 \Zcopemc \Big]~, \nonumber \\
& s^\y& \! \! \! \! \! (Y,\te) = - {1 \over 16 P^2 E}~\Big[ 3 E^2 \Zsipamc + (24 + 18 E^2) \Zsipame
- (12 - 21 E^2) \Zsipapa  \nonumber \\ &+& 12 E \Zsipbmd + 36 E \Zsipbnu
+ 2 E^2 \Zsipcmc  + (28 + 11 E^2) \Zsipcme  \nonumber \\
& + & 18 E \Zsipdmd + 3 E^2 \Zsipemc~\Big]~. \nonumber \feqq
The expressions of $v,p,q,w,u$ are too long to report all of them here. As
examples, we write down the components $v^\p$ and $u^\p$, which are
\beqq \! \! \! \! & v^\p& \!\!\!\!\!\!(I,\te) \! = \!
{1 \over 12 \pi P} \Big[ 16 E \sin Y - 18 E \sin(\te + Y) - 9 \sin(2 \te) - 2 E \sin(3 \te - Y)
\Big], \\
\! \! \! \! & u^\p& \!\!\!\!\!\!(I,\te) = {3 \pi \over 128 P^5} \Big[
- 512 E \Zsinupa + 824 E^2 \Zsinupb - 103 E^3 \Zsipamc \\
& + &  64 E^2 \Zsipamd - (1296 E - 36 E^3) \Zsipame
   - (768 - 1152 E^2) \Zsipanu   \nonumber \\
& + & (3524 E + 567 E^3) \Zsipapa + 960 E^2 \Zsipapb - 24 E^3 \Zsipapc   \nonumber \\
& - & 576 E^2 \Zsipbmd + 2048 E \Zsipbme + (4576 + 1992 E^2) \Zsipbnu    \nonumber \\
& + & 1024 E \Zsipbpa - 744 E^2 \Zsipbpb + 36 E^3 \Zsipcmc  \nonumber \\
& + & 1216 E^2 \Zsipcmd + (3180 E + 521 E^3) \Zsipcme - (1792 - 640 E^2) \Zsipcnu  \nonumber \\
& - & (1264 E + 172 E^3) \Zsipcpa - 27 E^3 \Zsipcpc
+ 560 E^2 \Zsipdmd   \nonumber \\
& - & 1536 E \Zsipdme + (256 - 912 E^2) \Zsipdnu - 336 E^2 \Zsipdpb  \nonumber \\
& + & 57 E^3 \Zsipemc - 320 E^2 \Zsipemd + (288 E - 144 E^3) \Zsipeme  \nonumber \\
& - & (900 E + 115 E^3) \Zsipepa + 88 E^2 \Zsiqa - (672 + 696 E^2) \Zsiqb  \nonumber \\
& + & 4 E^3 \Zsiqc - (812 E + 133 E^3) \Zsiqd - 328 E^2 \Zsiqe  \nonumber \\
& - & 45 E^3 \Zsiqf \Big]~. \nonumber \feqq
For future use, it is convenient to point out that
\beqq \overline{p}^{\, \p}(I) & = & - {3 \pi E^2 \over 2 P^3} \sin(2 Y)~, \label{compp1} \\
\overline{p}^{\, \e}(I) & = & {3 \pi \over 4 P^4} (10 E - E^3) \sin(2 Y)~, \nonumber \\
\overline{p}^{\, \y}(I) & = & {3 \pi \over 16 P^4} \Big[34 + 25 E^2 + (40 + 10 E^2) \cos(2 Y)\Big]~; \nonumber \feqq
\beq {\partial {\overline{f}}^{\y} \over \partial P}(I) = {6 \pi \over P^3}~, \qquad \FF^{i}_{\!j}(I)
= 0 \qquad \mbox{otherwise}~; \feq
\beq \MM^{i}_{j k}(I) = 0 \qquad \mbox{for all $i,j,k$}~. \feq
To go on, we take $\SS$ as in \rref{tays} and $\GG$, $\HH$ as in
\rref{tayf}; so,
\beq \hspace{-0.3cm} \SS^i(I,\delta I,\te)  = \!
\int_{0}^1 \! \! \! \! \! d \var \, {\partial s^{i} \over \partial I^j}(I+ \var \delta I, \te); \label{touses} \feq
\beq \hspace{-0.0cm} \GG^i_j(I, \delta I)  = \!
\int_{0}^1 \! \! \! d \var \, {\partial \overline{p}^{i}
\over \partial I^j} (I + \var \delta I); \label{touse} \feq
\beq \HH^{i}_{j k}(I, \delta I) = \int_{0}^1 d \var \, (1 - \var)
{\partial^2 \overline{f}^i \over \partial I^j \partial I^k}(I + \var \delta I)~, \label{he1} \feq
the derivatives of $s^i$, $\overline{f^i}$ and $\overline{p^i}$ being computed from Eqs.
\rref{sx} \rref{mf} and \rref{compp1}. In particular, we mention that
\beq {\partial^2 \overline{f}^\y \over \partial P^2}(I) = - {18 \pi \over P^4}~, \qquad
{\partial^2 \overline{f}^{i} \over \partial I^j \partial I^k}(I)= 0~~ \mbox{otherwise}. \feq
\textbf{\QC. The functions $\boma{\R, \K}$; time intervals.} These functions
are the solutions of Eqs. \rref{sistr} \rref{sistk},
and can be computed in an elementary way. The expressions
for $\R(\tau)$ and its inverse matrix are
\beq \R(\tau) =
\left( \barray{ccc} 1 & 0 & 0 \\ 0 & 1 & 0 \\
\\ \dd{6 \pi \over P^3_0} \, \tau & 0 & 1 \farray
\right)~; \qquad
\R^{-1}(\tau) = \left( \barray{ccc} 1 & 0 & 0 \\ 0 & 1 & 0 \\
\\ - \dd{6 \pi \over P^3_0} \, \tau & 0 & 1 \farray \right)~; \label{rrinv} \feq
concerning $\K$, we find
\beq\K^\p(\tau) =  {E^2_0 \over 4 P_0}~ \left[\cos(2 Y_0)
- \cos(2 Y_0 - {6 \pi \over P_0^2}\,\tau) \right]~, \label{kkx} \feq
$$ \K^\e(\tau)  = - {10 E_0 - E^3_0 \over 8 P_0^2} \left[\cos(2 Y_0)
- \cos(2 Y_0 - {6 \pi \over P_0^2}\,\tau) \right]~, $$
$$ \K^\y(\tau) = {3 \pi \over 16 P^4_0}~\Big[34 + 25 E^2_0 + 8 E^2_0 \cos(2 Y_0) \Big] \tau
+ {20 + E^2_0 \over 16 P^2_0} \left[\sin(2 Y_0) - \sin(2 Y_0 -  {6 \pi \over P_0^2}\, \tau) \right]~. $$
In the above, $\tau \in [0,+\infty)$; however, from now on we put a limitation
$\tau \in [0,U)$, for some finite $U$; consequently, the "orbit counter" $\ti$ will range in $[0,U/\ep)$.
\salto
\textbf{\QD. The auxiliary functions $\boma{\rho^i, \aiz, a^{i}_{j},}$
...,  $\boma{e^{i}_{j k}}$, $\boma{\alpha^i}$, $\boma{\gamma^i}$.}
These functions must be constructed so as to fulfill the inequalities in Section \ref{secav}.
First of all, we must find a function
$\rho = (\rho^i)  \in C([0,U), [0,+\infty]^3)$ \
such that $B(\J(\tau), \rho(\tau)) \subset \Lambda$ for all $\tau$, with $\Lambda$ as in Eq. \rref{delambda}.
The cited equation can be satisfied putting
\beq \rho^\p(\tau) := P_0~, \quad
\rho^\e(\tau) := \min(E_0, 1 - E_0)~, \qquad \rho^\y(\tau) := +\infty  \feq
for $\tau \in [0,U)$ (recall that $\J^\p(\tau) = P_0$ and $\J^\e(\tau) = E_0$).
From these functions,
we define $\Sgr$ as in Eq. \rref{dero}; this has elements $(\tau,r) = (\tau,r^\p,r^\e, r^\y)$. \parn
To go on, we must find a system of auxiliary functions
\beq \aiz \in C^{\infty}([0,U), [0,+\infty))~,
\quad a^{i}_{j}, b^i, c^i, d^{i}_{j}, e^{i}_{j k}  \in C^{\infty}(\Sgr, [0,+\infty)) \feq
such that Eqs. \rref{faz} \rref{faa} and \rref{fb}--\rref{fe} be satisfied. As explained
in subsection 2F, each function $\aiz$ should give an accurate upper bound for the left-hand side of Eq.
\rref{faz}; for this reason we have developed an algorithm based on numerical maximization over $\te$ and subsequent
interpolation in $\tau$, see Appendix \ref{appalg}. \par
Let us pass to the functions $a^{i}_{j}, b^i,...,e^{i}_{j k}$; we refer again to subsection 2F,
suggesting to find these functions by fairly rough majorizations of the left-hand sides
of Eqs. \rref{faa} and \rref{fb}--\rref{fe}. The expressions to be bounded are very lengthy in
some cases; they were majorized with the method described in Appendix \ref{appeae} (sometimes using
the symbolic mode of MATHEMATICA for the necessary computations).
Here, we only report the final expressions of the majorizing functions,
which are in fact $\tau$-independent; for this reason we write $a^{i}_{j}(r)$, $b^i(r)$, etc.,
instead of $a^{i}_{j}(\tau,r)$, $b^{i}(\tau,r)$, etc.. It should also be noted that the dependence on $r$
is through the components $r^\p$, $r^\e$. In the sequel, for the sake of brevity we put
\beq P_{\pm} := P_0 \pm r^\p~, \qquad E_{\pm} := E_0 \pm r^\e~; \feq
with these notations, we can take
\beqq a^\p_\p(r) \!\!\!\!&  := & {3 + 4 \Ep \over \Pm^2}~, \qquad\qquad a^\p_\e(r) :=  { 4 \over P_{-}}~,
\qquad \qquad a^\p_\y(r) :=  { 4 \Ep \over P_{-}}~, \label{axx} \\
a^\e_\p(r) \!\!\!\!& := & { 32 + 45 \Ep + 32 \Ep^2 \over 4 \Pm^3}~,
a^\e_\e(r) :=  { 45 + 64 \Ep \over  8 \Pm^2}~,
a^\e_\y(r) :=  {16 + 15 \Ep + 20 \Ep^2 \over 4 \Pm^2 }~,  \nonumber \\
a^\y_\p(r) \!\!\!\!& := & {32 + 33 \Ep + 29 \Ep^2 \over 4 \Pm^3 \Em}~,
a^\y_\e(r) :=  { 32 + 29 \Ep^2 \over 8 \Pm^2 \Em^2}~,
a^\y_\y(r) :=  {32 + 30 \Ep + 37 \Ep^2 \over 8 \Pm^2 \Em}~; \nonumber \feqq
\beqq b^\p(r) \!\!\!\!& := & \dd{ 54 + 112 \Ep + 33 \Ep^2 \over 8 \Pm^3 }~, \label{bx}  \\
b^\e(r) \!\!\!\!& := &  {6112 + 10832 \Ep + 6940 \Ep^2 + 11372 \Ep^3 +
    1441 \Ep^4 \over 512 \Pm^4 \Em}~, \nonumber \\
b^\y(r) \!\!\!\!& := & {1 \over 256 P_0^3 \Em^2 \Pm^4 }~
\Big[ 3520 P_0^3 + 16384 P_0^3 \Ep + 9340 P_0^3 \Ep^2  \nonumber \\
& + & 8940 P_0^3 \Ep^3 + 1861 P_0^3 \Ep^4 +
    1152 \Ep^2 \Pp^3 + 4608 \Ep^3 \Pp^3 \Big]~ ; \nonumber \\
c^\p(r) \!\!\!\!& := &  3 \pi {504 + 1024 \Ep + 713 \Ep^2 + 124 \Ep^3 \over 8 \Pm^5}~,
\label{cx} \\
c^\e(r) \!\!\!\!& := & {3 \pi \over 2048 \Pm^6 \Em^2}~
\Big[148736 + 738384 \Ep + 1062656 \Ep^2  \nonumber \\
& + & 1220344 \Ep^3 + 675146 \Ep^4 + 336591 \Ep^5 +
      26855 \Ep^6 \Big]~, \nonumber \\
c^\y(r)\!\!\!\! & := & {\pi \over 1024 P_0^3 \Em^3 \Pm^6}~
\Big[ 370944 P_0^3  + 2214336 P_0^3  \Ep + 5434752 P_0^3  \Ep^2  \nonumber \\
& + & 4927104 P_0^3  \Ep^3 + 2945040 P_0^3  \Ep^4 + 1225668 P_0^3  \Ep^5 + 147777 P_0^3  \Ep^6  \nonumber \\
& + &  231936  \Ep^3 \Pp^3 + 442368  \Ep^4 \Pp^3 +
    196608  \Ep^5 \Pp^3 \Big] ~; \nonumber \feqq
\beqq d^{\p}_{\p}(r) \!\!\!\!& := & {9 \pi \Ep^2 \over 2 \Pm^4}~, \qquad\qquad
d^{\p}_{\e}(r) := {3 \pi \Ep \over \Pm^3}~, \qquad \qquad
d^{\p}_{\y}(r) :=  {3 \pi \Ep^2 \over \Pm^3} ~, \label{dxx} \\
d^{\e}_{\p}(r) \!\!\!\!& := & {3 \pi \Ep (10 + \Ep^2 ) \over \Pm^5} ~,~
d^{\e}_{\e}(r) :=   {3 \pi (10 + 3 \Ep^2) \over 4 \Pm^4}~,~
d^{\e}_{\y}(r)  :=  {3 \pi \Ep (10 + \Ep^2) \over 2 \Pm^4} ~, \nonumber \\
d^{\y}_{\p}(r)\!\!\!\! & := & {3 \pi (74  + 35 \Ep^2) \over 4 \Pm^5}~,~
d^{\y}_{\e}(r) := {105 \pi \Ep \over 8 \Pm^4} ~, \quad
d^{\y}_{\y}(r) :=  {15 \pi (4 + \Ep^2) \over 4 \Pm^4}~; \nonumber \\
e^{\y}_{\p \p}(r) \!\!\!\!& := & {18 \pi \over \Pm^4}~, \qquad \qquad e^{i}_{j k}(\tau, r) := 0
\quad \mbox{otherwise}~. \label{eyh} \feqq
From the above functions we obtain a set of $C^{\infty}$ functions
$\alpha^i$, $\gamma^i$,
specializing the prescriptions \rref{al} \rref{ga} to the present case.
Explicitly,
\beq \alpha^i (\tau,r) := a^i(\tau, r) + \ep b^i(r) = \aiz(\tau) + a^{i}_{k}(r) r^k + \ep b^{i}(r)~,\label{ali} \feq
\beq \gamma^{i}(\tau,r,\ell) \equiv \gamma^i(r, \ell) := c^{i}(r) +
\d^{i}_{j}(r) \ell^\j + {1 \over 2} e^{i}_{j k}(r) \ell^j \ell^k~, \label{gai} \feq
for $(\tau,r) \in \Sgr$ and $\ell \in [0,+\infty)^3$.
\salto
\textbf{\QE. The matrix $\boma{(1 - \ep \partial \alpha/\partial r)}$ and its inverse.}
From Eq. \rref{ali}, one finds that the derivatives of $\alpha^i$ with respect to the $r$ variables
depend on $r$ but not on $\tau$. From the same equation, we see that our matrix has components
\beq \delta^i_j - \ep {\partial \alpha^i \over \partial r^j}(r) =
\delta^i_{j} - \ep \Big(\,{\partial a^{i}_{k} \over \partial r^j}(r) r^k +
a^{i}_{j}(r) + \ep {\partial b^i \over \partial r^j}(r)\,\Big), \qquad (i,j=\p, \e, \y)~; \label{matel} \feq
all the derivatives in the right hand side are easily obtained from Eq.s \rref{axx} \rref{bx}.
Eq. \rref{quattro}, that is basic in our approach, contains the inverse of this matrix;
this could be written explicitly, but
its expression is uselessly complicated. For this reason, in the subsequent numerical computations we
use an approximation giving the inverse up to a third order error
$O_3(\ep, r)$, for $(\ep,r) = (\ep, r^p, r^\e, r^\y) \vain 0$
({\footnote{Recall that, in Eq. \rref{quattro}, the matrix in which we are interested
is evaluated with $r^i = \ep \en^i(\tau)$; with this position for $r$, \rref{invapp} gives
an inverse up to $O(\ep^3)$.}}). Neglecting this error, we have
\beq \Big(1 - \ep {\partial \alpha \over \partial r}(r)\Big)^{-1} \!\!= 1 + \ep M +
\ep r^k N_{(k)} + \ep^2 \Q~, \label{invapp} \feq
(with a sum for $k \in \{\p,\e,\y\}$), where we have introduced the matrices
\beq M := \left( \barray{ccc} \dd{3 + 4 E_0 \over P_0^2} & \dd{4 \over P_0} & \dd{4 E_0 \over P_0} \\
\dd{32 + 45 E_0 + 32 E_0^2 \over 4 P_0^3} & \dd{45 + 64 E_0 \over 8 P_0^2} & \dd{16 + 15 E_0 + 20 E_0^2 \over 4 P_0^2} \\
\dd{32 + 33 E_0 + 29 E_0^2 \over 4 E_0 P_0^3} & \dd{32 + 29 E_0^2 \over 8 E_0^2 P_0^2} &
\dd{32 + 30 E_0 + 37 E_0^2 \over 8 E_0 P_0^2} \farray \right)~; \label{emme} \feq
\beqq N_{(\p)} \!\!\!& := & \left( \barray{ccc} \dd{4(3 + 4 E_0) \over P_0^3} & \dd{8 \over P_0^2} &
\dd{4 E_0 \over P_0^2} \\ \dd{3(32 + 45 E_0 + 32 E_0^2) \over 2 P_0^4} &
\dd{45 + 64 E_0 \over 2 P_0^3} & \dd{16 + 15 E_0 + 20 E_0^2 \over 2 P_0^3} \\
\dd{3(32 + 33 E_0 + 29 E_0^2) \over 2 E_0 P_0^4} & \dd{32 + 33 E_0 + 58 E_0^2 \over 2 E_0^2 P_0^3} &
\dd{32 + 30 E_0 + 37 E_0^2 \over 4 E_0 P_0^3} \farray \right)~, \nonumber \\
N_{(\e)} \!\!\! &:= & \left( \barray{ccc} \dd{8 \over P_0^2} & \dd{0} &
\dd{4 \over P_0} \\ \dd{45 + 64 E_0 \over 2 P_0^3} &
\dd{16 \over P_0^2} & \dd{5(3 + 8 E_0) \over 4 P_0^2} \\
\dd{32 + 33 E_0 + 58 E_0^2 \over 2 E_0^2 P_0^3} & \dd{16 + 29 E_0^2 \over E_0^3 P_0^2} &
\dd{32 + 60 E_0 + 111 E_0^2 \over 8 E_0^2 P_0^2} \farray \right)~, \nonumber \\
N_{(\y)} \!\!\! & := & \left( \barray{ccc} \dd{4 E_0 \over P_0^2} & \dd{4 \over P_0} &
\dd{0} \\ \dd{16 + 15 E_0 + 20 E_0^2 \over 2 P_0^3} &
\dd{5(3 + 8 E_0) \over 4 P_0^2} & \dd{0} \\
\dd{32 + 30 E_0 + 37 E_0^2 \over 4 E_0 P_0^3} & \dd{32 + 60 E_0 + 111 E_0^2 \over 8 E_0^2 P_0^2} &
\dd{0} \farray \right)~; \label{enne}
\feqq
$$ \Q := \left( \barray{ccc}
\dd{ 746 + 1152 E_0 + 715 E_0^2 \over 8 P_0^4} \vspace{0.2cm}
& \dd{ 64 + 194 E_0 + 283 E_0^2 \over 4 E_0 P_0^3}
& \dd{ 64 + 84 E_0 + 109 E_0^2\over 2 P_0^3} \\
\dd{ \QQ^{\e}_{\p} \over 128 E_0 P_0^5} \vspace{0.2cm}
& \dd{ \QQ^{\e}_{\e} \over 512 E_0^2 P_0^4}
& \dd{ \QQ^{\e}_{\y} \over 32 E_0 P_0^4} \\
\dd{ \QQ^{\y}_{\p} \over 64 E_0^2 P_0^5}
& \dd{ \QQ^{\y}_{\e} \over 128 E_0^3 P_0^4}
& \dd{ \QQ^{\y}_{\y} \over 64 E_0^2 P_0^4}
\farray \right)~, $$
\beqq \QQ^{\e}_{\p} \!\!\! & := & 10208 + 27728 E_0 + 44440 E_0^2 + 46244 E_0^3 + 18369 E_0^4~, \nonumber \\
\QQ^{\e}_{\e} \!\!\! & := & 14304 + 29344 E_0 + 71068 E_0^2 + 121568 E_0^3 + 65637 E_0^4~, \nonumber \\
\QQ^{\e}_{\y} \!\!\! & := & 512 + 1680 E_0 + 4405 E_0^2 + 4455 E_0^3 + 3044 E_0^4~, \nonumber \\
\QQ^{\y}_{\p} \!\!\! & := & 7616 + 24832 E_0 + 25096 E_0^2 + 27300 E_0^3 + 7719 E_0^4~, \nonumber \\
\QQ^{\y}_{\e} \!\!\! & := & 5568 + 29376 E_0 + 33400 E_0^2 + 42444 E_0^3 + 15153 E_0^4~, \nonumber \\
\QQ^{\y}_{\y} \!\!\! & := & 2048 + 2880 E_0 + 7524 E_0^2 + 5202 E_0^3 + 4385 E_0^4~. \label{qu}
\feqq
For more details on the approximate inverse \rref{invapp}, see Appendix \ref{appinv}.
\salto
\textbf{\QF. The functions $\boma{R^{i}_{j}}$, $\boma{P^{i}_{j}}$.} According to
Eq. \rref{rp}, these should fulfill the inequalities
\beq | \R(\tau) |^{i}_{j} \leqs R^{i}_{j}(\tau)~,~~
| \R^{-1}(\tau) |^{i}_{j} \leqs P^{i}_{j}(\tau) \quad \mbox{for
$\tau \in [0,U)$}~. \feq
On account of Eq. \rref{rrinv}, we can take
\beq \big(R^{i}_{j}(\tau)\big) := \big(P^{i}_{j}(\tau)\big) :=
\left( \barray{ccc} 1 & 0 & 0 \\ 0 & 1 & 0 \\
\\ \dd{6 \pi \over P^3_0} \, \tau & 0 & 1 \farray \right)~.
\feq
\salto
\textbf{\QG. The main result: the estimates $\boma{| L^i(t) | \leqs \en^i(\ep t)}$
from Proposition \ref{proprinc}. The ``$\boma{\Np}$-operation''.} We are finally ready to discuss the difference
\beq \L(t) := {1 \over \ep}~[\I(t) - \J(\ep t)] =
{1 \over \ep} \Big(\P(t) - P_0, \E(t) - E_0, \Y(t) - \J^\y(\ep t) \Big) \label{diff} \feq
where $\I = (\P,\E, \Y)$ is the solution of Eqs. \rref{eqpey}.
We follow the scheme of Proposition \ref{proprinc}; this requires
to determine $\ell_0 = (\ell^\p_0, \ell^\e_0, \ell^\y_0)$ by solving a
fixed point problem
\beq \alpha^i(0, \ep \ell_0) = \ell^i_0~. \feq
In the examples that follow $\ell_0$ is always found numerically, as explained in
Remark \ref{rema2}(a).
After $\ell_0$ has been found,  we must solve (again numerically) a Cauchy problem for the unknown functions
$\em = (\em^i), \en = (\en^i) \in C^1([0,U),\reali^3)$, i.e.,
\beq {d \em^i \over d \tau} =
P^{i}_{j} \, \gamma^j(\ep \en, \en)~, \qquad
\em^i(0) = 0~, \label{trei} \feq
\beq {d \en^i \over d \tau} = \Big(1 - \ep {\partial \alpha \over
\partial r}\, (\cdot, \ep \en)
\Big)^{-1, i}_{~~~ k}
\left( {\partial \alpha^k \over \partial \tau}\,(\cdot , \ep \en)
+ \ep R^{k}_{h} P^{h}_{j}
\, \gamma^j(\ep \en, \en) + \ep {d R^{k}_{j} \over d \tau} \em^j \right)~,
\label{quattroi} \feq
$$ \qquad \en^i(0) = \ell^i_0  $$
with the domain conditions \rref{due}, having in this case the form
$$ 0 < \ep \en^\p < P_0~, ~~ 0 < \ep \en^\e < \min(E_0, 1 - E_0)~, ~~\en^\y > 0~, $$
\beq \det \Big(1 - \ep {\partial \alpha \over \partial r}\,(\cdot, \ep \en)
\Big) > 0~. \label{duei} \feq
If the above problem has solution on $[0,U)$, our general framework grants that the solution
$\I = (\P,\E,\Y)$ of \rref{eqpey} exists on $[0,U/\ep)$, and gives
the bounds
\beq | \L^i(t) | \leqs \en^i(\ep t) \qquad \mbox{for $t \in [0,U/\ep)$}~, \feq
i.e.,
\beq | \,\P(t) - P_0 | \leqs \ep \en^\p(\ep t), ~~
| \, \E(t) - E_0 | \leqs \ep \en^\e(\ep t),~~
| \, \Y(t) - \J^\y(\ep t) | \leqs \ep \en^\y(\ep t). \feq
With a slight variation of the terminology of \cite{uno} \cite{due}, we
call ``$\Np$-operation'' the execution of \textsl{all} the numerical computations required by
the present approach to obtain
the estimators $(\en^i)$,
given the initial data $P_0, E_0, E_0$ and the values of $\ep, U$. These computations include: \par \noindent
(a) the numerical evaluations and subsequent interpolations, necessary to determine the functions
$\aiz$ $(i=\p,\e,\y)$ on $[0,U)$. We have already mentioned Appendix \ref{appalg}, that describes
everything in detail; \par \noindent
(b) the determination of the fixed point $\ell_0$; \par \noindent
(c) the numerical solution of the Cauchy problem \rref{trei} \rref{quattroi} for $(\em^i)$, $(\en^i)$.
\par
From now on, we will indicate with $\TT_{\Np}$ the CPU time required to perform (a)(b)(c);
this time depends on the chosen  hardware and software, but is anyhow useful for comparison with other
computational approaches, such as the one described hereafter.
\salto
\textbf{\QH. The ``$\Lp$-operation'': a way to
test the efficiency of the previous estimates.}
This operation was already considered in \cite{due}, with a
role similar to the one that we ascribe to it here: checking the reliability of the $\Np$- estimates
by a direct numerical treatment of the system \rref{eqpey}, for $t \in [0,U/\ep)$.
The direct attack of \rref{eqpey} is possible when the time scale $U/\ep$ is not overwhelmingly large;
for larger $U/\ep$, the $\Lp$-operation is too expensive, and this is just
the case where the $\Np$- procedure becomes more useful. \par
The direct attack to \rref{eqpey} is performed substituting in these
equations $\P(t) = P_0 + \ep \L^\p(t)$, $\E(t) = E_0 + \ep \L^\e(t)$,
$\Y(t)  = \J^\y(\ep t) + \ep \L^\y(t)$, which gives rise to the equations
\beq {d \L^\p \over d t}(t) = f^\p(P_0 + \ep \L^\p(t), E_0 + \ep \L^\e(t),
\J^\y(\ep t) + \ep \L^\y(t)),~ \label{pertel} \feq
$$ {d \L^\e \over d t} (t) = f^\e(P_0 + \ep \L^\p(t), E_0 + \ep \L^\e(t),
\J^\y(\ep t) + \ep \L^\y(t)),~$$
$$ {d \L^\y \over d t}(t) = f^\y(P_0 + \ep \L^\p(t), E_0 + \ep \L^\e(t),
\J^\y(\ep t) + \ep \L^\y(t)) + {3 \pi \over \J^\p(\ep t)^2}~, $$
$$ (\L^\p, \L^\e, \L^\y)(0) = 0~. $$
By definition, the $\Lp$-operation is the numerical solution of this Cauchy problem
in the unknowns $\L^i$, for $t \in [0,U/\ep)$; $\TT_{\Lp}$ will indicate the necessary
CPU time.
\salto
\section{Examples.}
\label{seces}
\textbf{Introducing the examples.}
In the case of the Earth,
\beq G M = 3.98600442 \cdot 10^{14} {\mbox{m}^3 \over \mbox{sec}^2}~,
\qquad R = 6.378135 \cdot 10^6 \, \mbox{m}~, \feq
$$ \ep = 5.457 \cdot 10^{-4}~. $$
So, for a satellite on a polar orbit around the Earth, the unperturbed apogee and perigee $\rho_{+}$,
$\rho_{-}$ and the orbital period $T_{orb}$ are determined in this way by the initial data $(P_0, E_0, Y_0)$:
$$ \rho_{\mp} = {P_0 \over 1 \pm E } \times 6.378135 \times 10^3 \, \mbox{Km},~~
T_{orb} = {P_0^{3/2} \over (1 - E^2_0)^{3/2}} \times 1.408150 \, \mbox{hours}. $$
In the sequel we present two examples, with initial data $(P_0, E_0, Y_0)$ very
similar to the actual data of two real satellites
(Polar and Cos-B). The MATHEMATICA package, already mentioned in
relation to the symbolic treatment of the problem, has been employed
(on a PC) for the required numerical computations. \parn
We repeat a comment of the Introduction: the examples are not fully realistic
since they do not account for perturbations different from the $J_2$ gravitational term; nevertheless,
we think they have some interest because they show the effectiveness of the method, suggesting
that our approach could be applied as well including other perturbations. \parn
Each example is worked out along these lines: \par \noindent
(i) Firstly, we take $U/\ep = 3000$. We perform both the $\Np$- and the $\Lp$-operations,
using the second to test the first one. We give figures reporting the
graphs of $\ep |\, \L^i(t)|$ (rapidly oscillating) and $\ep \en^i(\ep t)$, for $i \in \{\p, \e, \y\}$
and $t \in [0,U/\ep)$; the CPU times $\TT_{\Np}$, $\TT_{\Lp}$ are also indicated.
\par \noindent
(ii) Next, we choose $U/\ep = 60000$.
The $\Np$-operation is still performed within short CPU
times; we give figures reporting $\TT_{\Np}$ and the graphs of the estimators $(\ep \en^i)$.
On the contrary, the $\Lp$-operation exceeds the capabilities of the machine employed. \par
\salto
\textbf{Example 1.} We take
\beq P_0 := 3.000~, \qquad E_0 := 0.6640~, \qquad Y_0 := 0.0000~, \feq
corresponding to
\beq \rho_{+} = 56950 \, \mbox{Km}~, \qquad \rho_{-} = 11500 \, \mbox{Km}~, \qquad
T_{orb} = 17.50 \, \mbox{hours}~. \feq
These data are very similar to the initial conditions of the Polar satellite, taking $t=0$ on April 1997
\cite{Pol}. Eq. \rref{solmed} for the solution of the averaged system on any interval $[0,U)$ gives
$\J^\p(\tau) = \mbox{const.} = P_0$, $\J^\e(\tau) = \mbox{const.} = E_0$, and
\beq \J^\y(\tau) = - 1.047 \tau~. \feq
Figures 1a, 1b and 1c report the graphs of $\en^i(\ep t)$ and $| \L^i(t) |$ for
$t \in [0,3000]$, as produced by the $\Lp$- and $\Np$-operations. We note
that $3000 \, T_{orb} \simeq 6$ years.
The CPU times (in seconds) for $\Np$ and $\Lp$ reported in Fig.1a refer
to the overall calculation, also including the $\e$ and $\y$ components plotted in Figures 1b, 1c. \parn
Figures 1d, 1e and 1f report the graphs of $\en^i(\ep t)$ and $| \, \L^i(t) |$ for
$t \in [0,60000]$, as produced by the $\Np$-operations. We note that
$60000 \, T_{orb} \simeq 120$ years.
The CPU time indicated in Fig.1d also includes
calculations of the $\e$ and $\y$ components, plotted in Figures 1e, 1f.
In comparison with the previous case $t \in [0,3000]$,
the number of orbits is increased by a factor $20$;
in spite of this, $T_{\Np}$ changes by a factor $\lesssim 2$. \par
\vskip 0.2cm \noindent
\textbf{Example 2.} We take
\beq P_0 := 1.973~, \qquad E_0 = 0.8817~, \qquad Y_0 := 0.9600~, \feq
corresponding to
\beq \rho_{+} = 106400 \, \mbox{Km}~, \qquad \rho_{-} = 6688 \, \mbox{Km}~, \qquad
T_{orb} = 37.16 \, \mbox{hours}~. \feq
These data are very similar to the initial conditions at launch (August 1975)
of the Cos-B satellite \cite{Cosb}.
Eq. \rref{solmed} for the solution of the averaged system on any interval $[0,U)$ gives
$\J^\p(\tau) = \mbox{const.} = P_0$~, $\J^\e(\tau) = \mbox{const.} = E_0$ and
\beq \J^\y(\tau) = 0.9600 - 2.421 \, \tau~. \feq
All figures for this example give the same information as the corresponding ones of Example 1.
In particular, Figures 2a, 2b and 2c refer to computations for $t \in [0,3000]$, while Figures
2d, 2e, 2f refer to the case $t \in [0, 60000]$.
We note that $3000 \, T_{orb} \simeq 12$ years and
$60000 \, T_{orb} \simeq 250$ years. The CPU times (in seconds) for $\Np$ and $\Lp$ are similar to the
ones in Example 1, so the comments given therein also apply to this case.
\vfill
\begin{figure}
\parbox{3in}{
\includegraphics[
height=2.0in,
width=2.8in
]%
{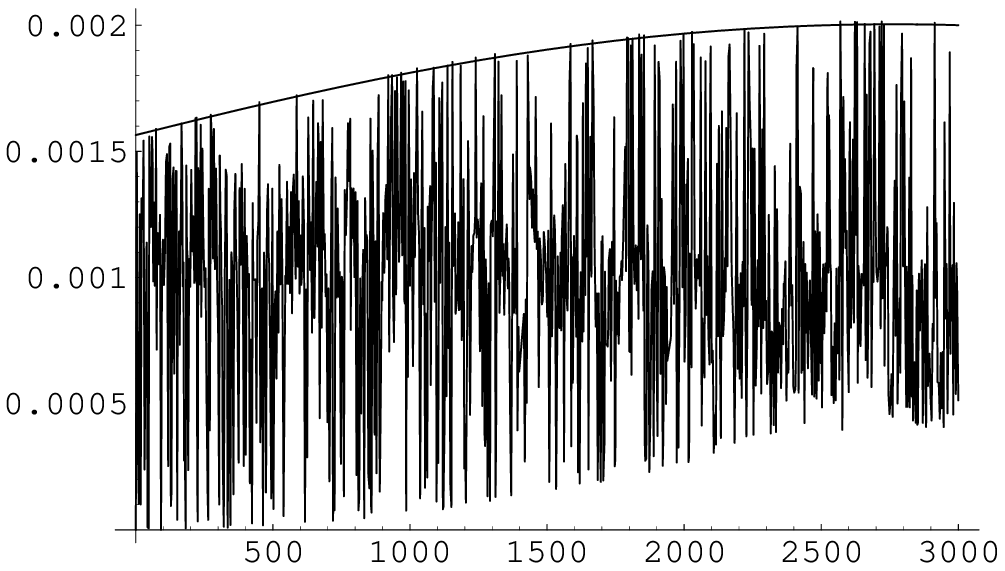}%
\parn
{\textbf{Figure 1a.~} Example 1, for $t \in [0,3000]$. Total CPU times:
$T_{\Np} = 5.18$ sec, $T_{\Lp} = 29.08$ sec.
Graphs of $\ep \en^\p(\ep t)$ and $\ep |\L^\p(t)|$. \parn}
\label{f1a}
}
\hskip 0.4cm
\parbox{3in}{
\includegraphics[
height=2.0in,
width=2.8in
]%
{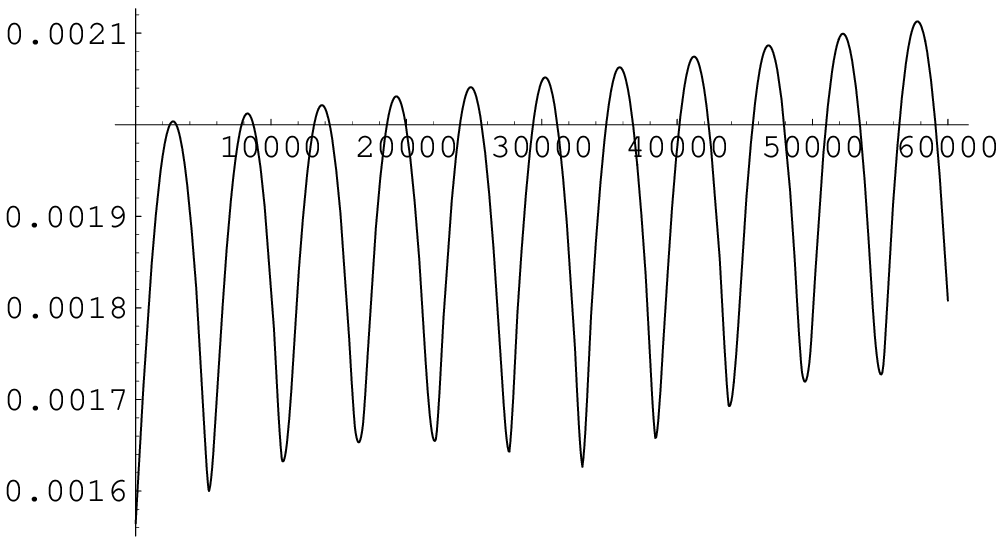}%
\parn
{\textbf{Figure 1d.~} Example 1, for $t \in [0,60000]$. Total CPU time:
$T_{\Np} = 10.11$ sec.
Graph of $\ep \en^\p(\ep t)$. \parn}
\label{f1d}
}
\parbox{3in}{
\includegraphics[
height=2.0in,
width=2.8in
]%
{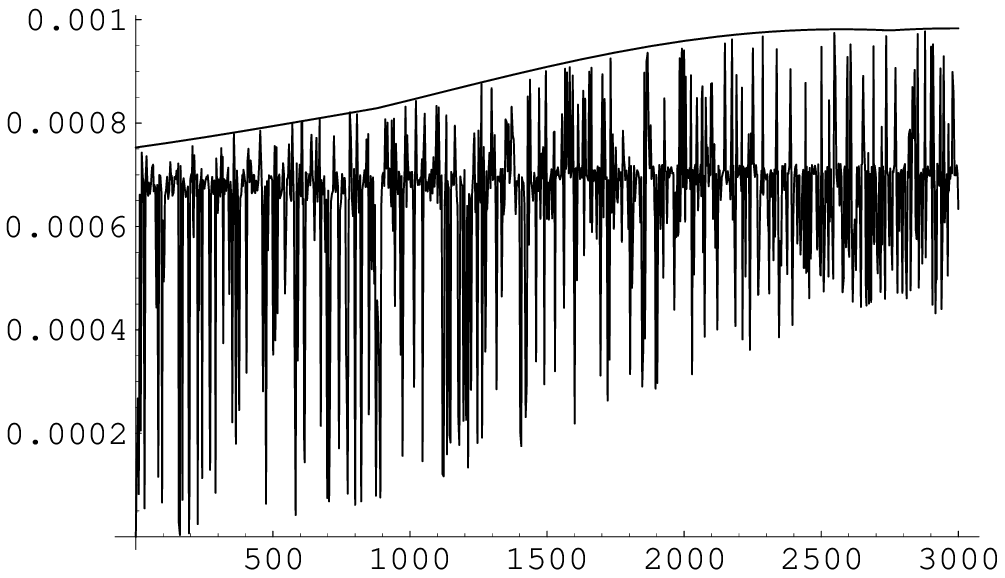}%
\parn
{\textbf{Figure 1b.~} The same situation as in Fig. 1a.
Graphs of $\ep \en^\e(\ep t)$ and $\ep |\L^\e(t)|$. \parn}
\label{f1b}
}
\hskip 0.4cm
\parbox{3in}{
\includegraphics[
height=2.0in,
width=2.8in
]%
{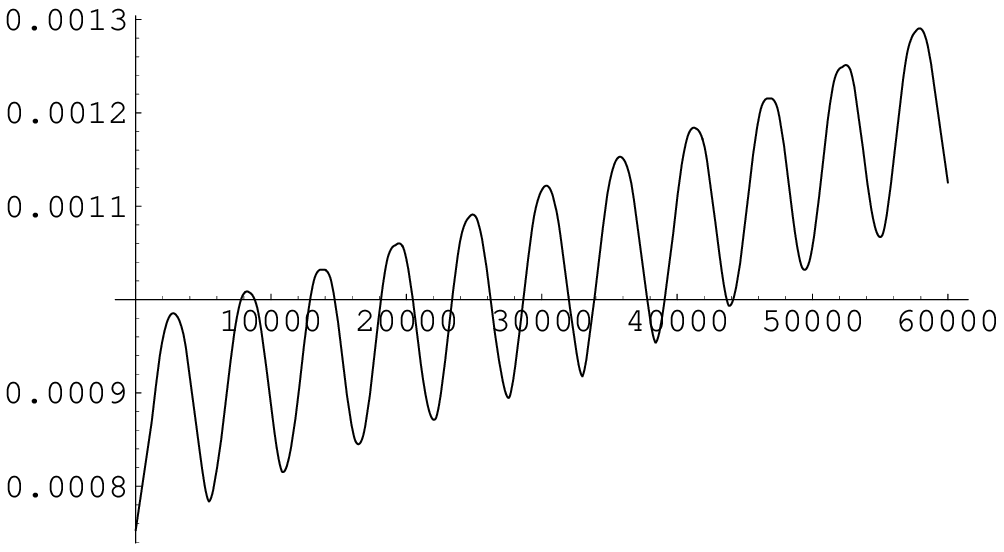}%
\parn
{\textbf{Figure 1e.~} The same situation as in Fig. 1d.
Graph of $\ep \en^\e(\ep t)$. \parn}
\label{f1e}
}
\parbox{3in}{
\includegraphics[
height=2.0in,
width=2.8in
]%
{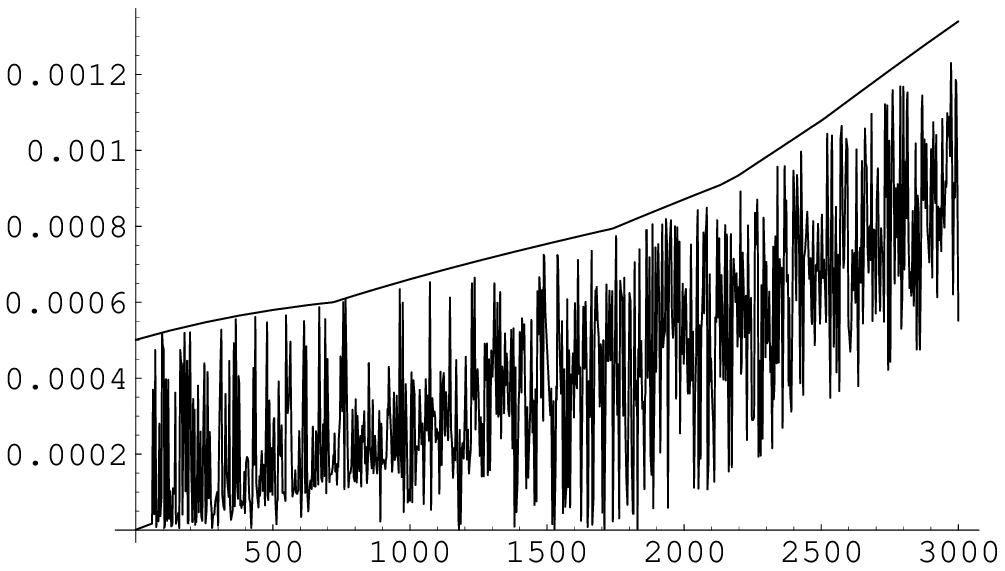}%
\parn
{\textbf{Figure 1c.~} The same situation as in Fig. 1a.
Graphs of $\ep \en^\y(\ep t)$ and $\ep | \L^\y(t) |$. \parn~\parn}
\label{f1c}
}
\hskip 0.4cm
\parbox{3in}{
\includegraphics[
height=2.0in,
width=2.8in
]%
{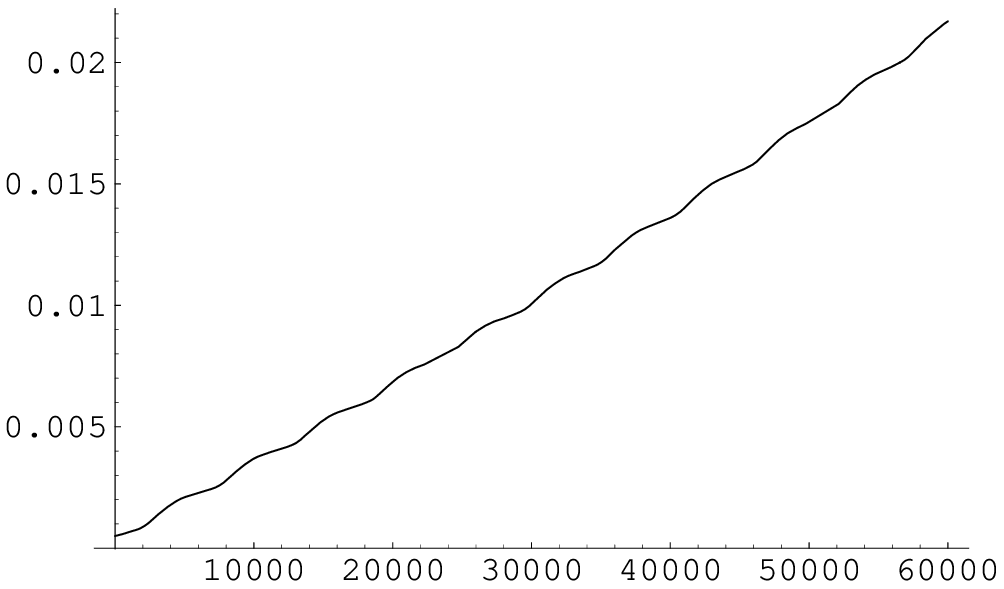}%
\parn
{\textbf{Figure 1f.~} The same situation as in Fig. 1d.
Graph of $\ep \en^\y(\ep t)$.
(Note that $\J^\y(\ep t) = - 34.29$ for $t=60000$).
\parn}
\label{f1f}
}
\end{figure}
\begin{figure}
\parbox{3in}{
\includegraphics[
height=2.0in,
width=2.8in
]%
{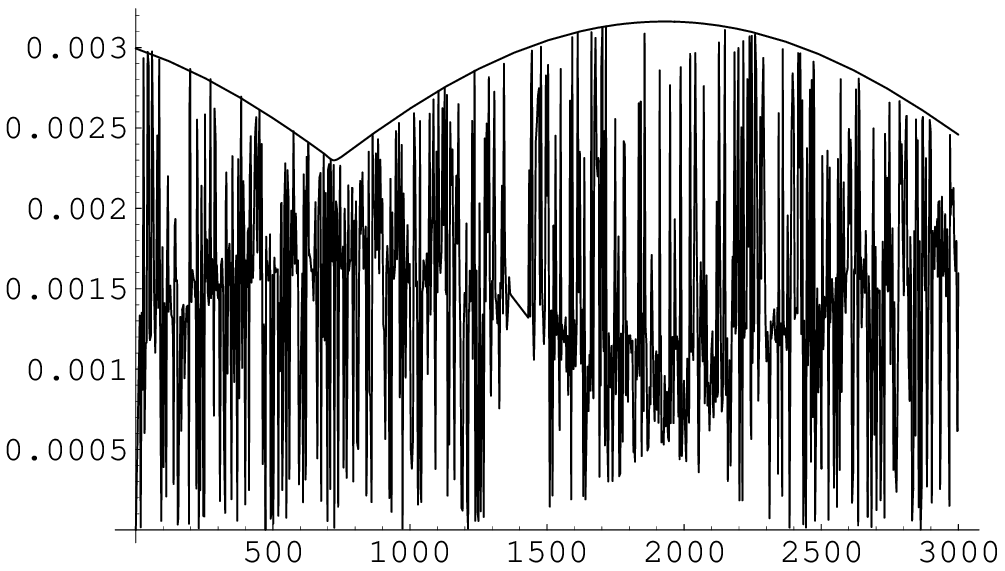}%
\parn
{\textbf{Figure 2a.~} Example 2, for $t \in [0,3000]$. Total CPU times:
$T_{\Np} = 5.77$ sec, $T_{\Lp} = 33.02$ sec.
Graphs of $\ep \en^\p(\ep t)$ and $\ep |\L^\p(t)|$. \parn}
\label{f2a}
}
\hskip 0.4cm
\parbox{3in}{
\includegraphics[
height=2.0in,
width=2.8in
]%
{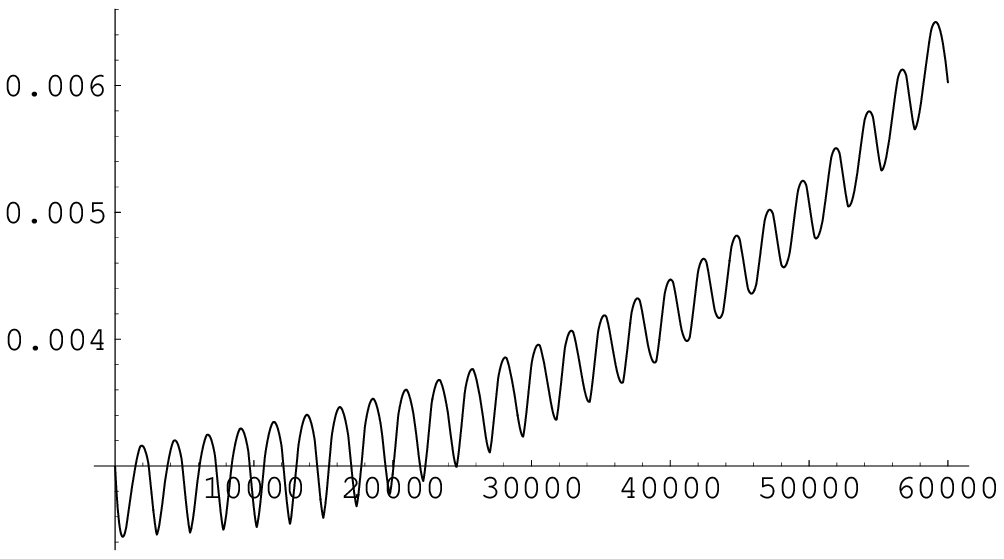}%
\parn
{\textbf{Figure 2d.~} Example 2, for $t \in [0,60000]$. Total CPU time:
$T_{\Np} = 11.28$ sec.
Graph of $\ep \en^\p(\ep t)$. \parn}
\label{f2d}
}
\parbox{3in}{
\includegraphics[
height=2.0in,
width=2.8in
]%
{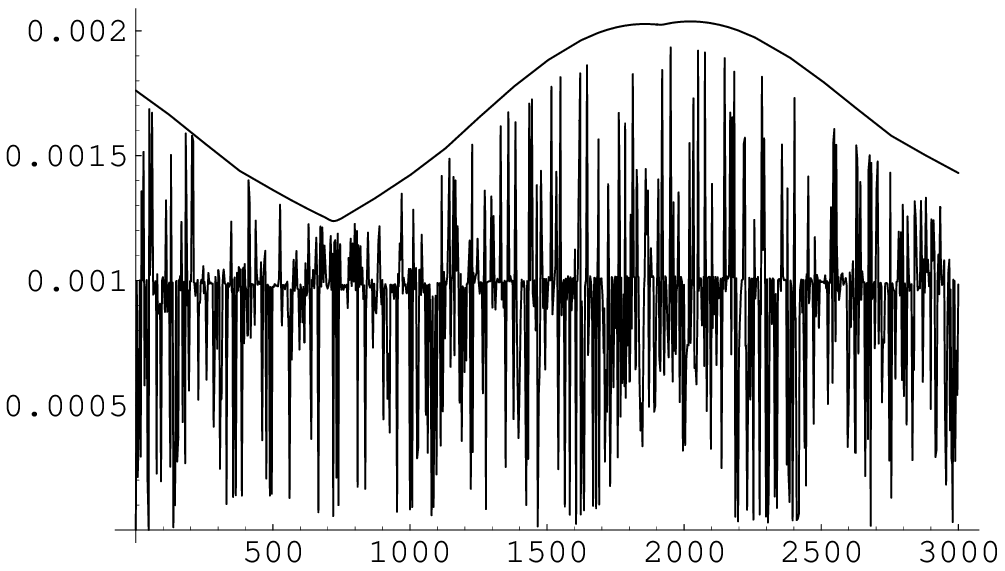}%
\parn
{\textbf{Figure 2b.~} The same situation as in Fig. 2a.
Graphs of $\ep \en^\e(\ep t)$ and $\ep |\L^\e(t)|$. \parn}
\label{f2b}
}
\hskip 0.4cm
\parbox{3in}{
\includegraphics[
height=2.0in,
width=2.8in
]%
{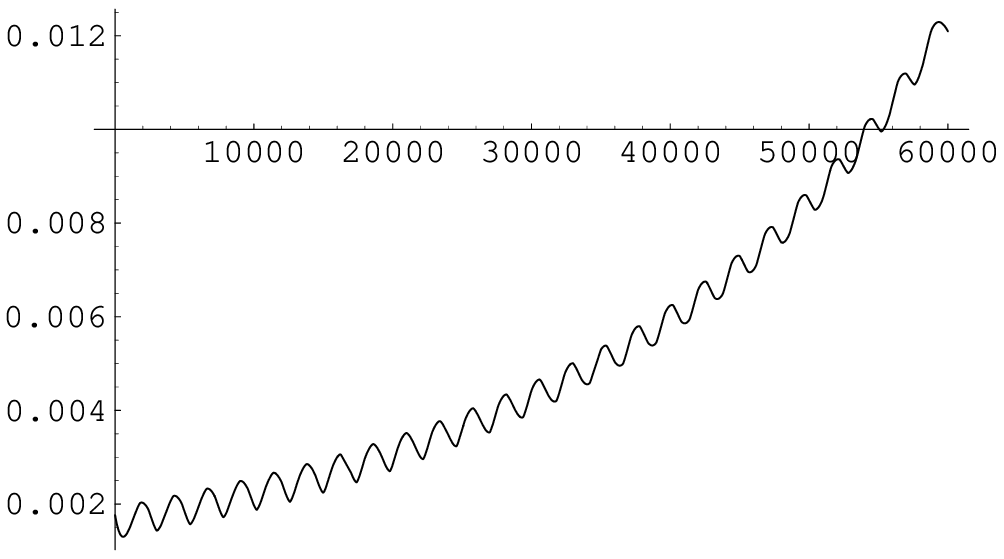}%
\parn
{\textbf{Figure 2e.~} The same situation as in Fig. 2d.
Graph of $\ep \en^\e(\ep t)$. \parn}
\label{f2e}
}
\parbox{3in}{
\includegraphics[
height=2.0in,
width=2.8in
]%
{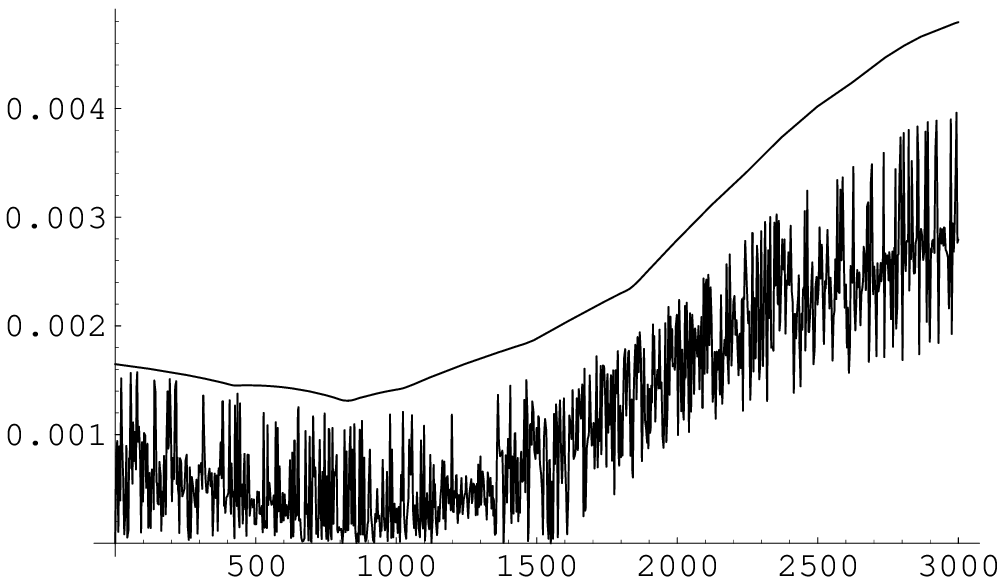}%
\parn
{\textbf{Figure 2c.~} The same situation as in Fig. 2a.
Graphs of $\ep \en^\y(\ep t)$ and $\ep | \L^\y(\ep t) |$. \parn ~ \parn}
\label{f2c}
}
\hskip 0.4cm
\parbox{3in}{
\includegraphics[
height=2.0in,
width=2.8in
]%
{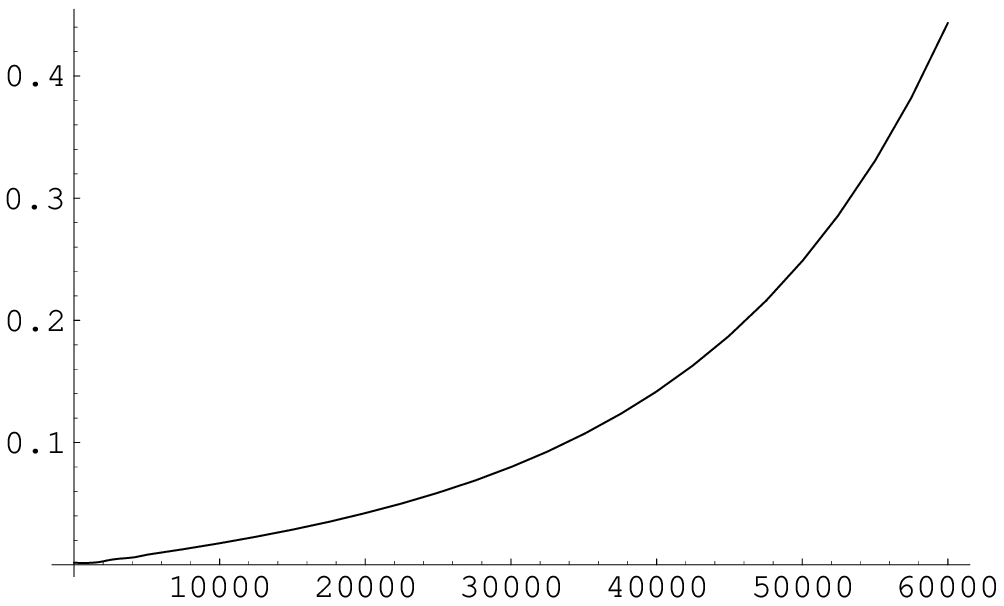}%
\parn
{\textbf{Figure 2f.~} The same situation as in Fig. 2d.
Graph of $\ep \en^\y(\ep t)$. (Note that $\J^\y(\ep t) = -78.31$ for $t=60000$)
\parn}
\label{f2f}
}
\end{figure}
\vfill \eject \noindent
\appendix
\section{Appendix. On Kepler elements.}
\label{appkepl}
We refer to the framework of subsection \satb, having fixed $(\X, \te) = (\rot,\tet,\ro, \te) \in \DD$.
The unperturbed Kepler orbit $\oxt$
with these initial data at time $t=0$ is an ellipse with (dimensionless) parameter
$P \equiv P(\X,\te)$, eccentricity $E \equiv E(\X,\te)$
and argument of the pericenter $Y \equiv Y(\X, \te)$.
In any textbook on classical mechanics, one finds the equations
\beq \ro = {R P \over 1 + E \cos(\te - Y)}~,\label{conics0} \feq
\beq \rot = { R P E \sin(\te - Y) \over (1 + E \cos(\te - Y))^2} \tet~, \label{conics} \feq
\beq  R P =  { \ro^4 \tet^2 \over G M}~, \label{para} \feq
\beq E = \sqrt{1 + {2  \ro^4 \tet^2 \over G^2 M^2}
\left( {1 \over 2} \rot^2 + {1 \over 2} \ro^2 \tet^2 - {G M  \over \rho}
\right)}~. \label{ecc} \feq
Eq. \rref{conics0} is the representation of a conic in polar coordinates;
Eq. \rref{conics} follows taking the derivative of this polar representation
with respect to time along the unperturbed Kepler motion, i.e.,
regarding $P,E,Y$ as time independent. Eqs. \rref{para} \rref{ecc} give
the parameter and the eccentricity as functions of the angular
momentum and the total energy (per unit mass) $\ro^2 \tet$ and
$(1/2)  \rot^2 + (1/2) \ro^2 \tet^2 - G M /\rho$~. \par
We note that Eq. \rref{conics0} and Eqs. \rref{para} \rref{ecc} are
equivalent to Eq. \rref{conics0}, to the square of Eq.\rref{conics} and to Eq. \rref{para}.
Eq. \rref{conics} is important because (due to $\tet > 0$ in $\DD$) it implies
\beq \sign \rot =  \sign \sin(\te - Y)~; \label{sign}\feq
this information is essential to determine $Y$ as a function of $(\xi,\te)$.
Eqs. \rref{conics0}-\rref{sign} yield the expressions for $P,E,Y$ in Eqs. \rref{yie1}-\rref{yie3};
from here, one also proves the bijectivity of the map $(\xi,\te)
\mapsto (P,E,Y,\te) \equiv (I,\te)$ and the expression \rref{inv0}-\rref{inv1} for its inverse.
\par
To conclude, let us comment on the equations \rref{torb} for the apocenter $\rho_{+}$, the
pericenter $\rho_{-}$ and the orbital period $T_{orb}$ of the unperturbed Kepler motion along
an ellipse of parameter $P$ and eccentricity $E$. The expressions for
$\rho_{\pm}$ are consequences of the polar equation for the ellipse;
the expression for $T_{orb}$ follows from Kepler's third law
\beq T_{orb} = {2 \pi \over \sqrt{G M}}~ (\mbox{semi-major axis})^{3/2} =
{2 \pi \over \sqrt{G M}} \left( {\rho_{+} \over 2} + {\rho_{-} \over 2} \right)^{3/2} \feq
and from the previously mentioned formulas for $\rho_{\pm}$.
\section{Appendix. Numerical computation of the functions $\boma{\aiz}$.}
\label{appalg}
Let us consider a general periodic system \rref{evol}, and suppose we have
the related functions $s,\J,\R,\K$, for $\tau$ within a finite interval $[0,U)$. Hereafter we
outline a simple, but effective scheme to construct by mixed, numerical and analytical,
techniques a function $\tau \in [0,U) \mapsto \aiz(\tau)$
fulfilling \textsl{up to small errors} the inequality \rref{faz}, for
any given $i \in \{1,...,d\}$; this method is easily implemented
on MATHEMATICA. The scheme consists of four items. \par \noindent
(i) Take on the torus $\bT$ a grid of equally spaced angles
\beq \te_q := \big[{2 \pi q \over Q}\big]~, \qquad (q = 1,2,...,Q)~. \label{teq} \feq
(ii) Take in $[0,U)$ a grid of equally spaced instants
\beq \tau_n := {U \over N}\, n~, \qquad n = 0,1,...,N-1~. \label{taun} \feq
(iii) For each $n$, compute numerically
\beq \aizn := \max_{q=1,...,Q} \left|
\Big( s(\J(\tau_n),\te_q) - \R(\tau_n) s(I_0, \te_0) - \K(\tau_n) \Big)^i \right|~; \label{ain} \feq
then, for all $\te \in \bT$,
\beq \left| \Big( s(\J(\tau_n),\te) - \R(\tau) s(I_0, \te_0) - \K(\tau_n) \Big)^i \right| \leqs \aizn \feq
up to an error neglected in the sequel, which is of order $O(1/Q^2)$ for
$Q \vain +\infty$. \par \noindent
(iv) Interpolate the sequence $a^i_{n}$ ($n=0,...,N)$, finding a
smooth function $\tau \in [0,U) \mapsto \aiz(\tau)$ such that
\beq \aiz(\tau_n) = \aizn \qquad \mbox{for~~ $n=0,1,...,N-1$.} \feq
More precisely, we define $\aiz$ to be the Lagrange
polynomial (restricted to $[0,U)$) such that $\aiz(\tau_n) = \aizn$ for all $n$; thus
\beq \aiz(\tau) = \sum_{n=0}^{N-1} \aizn~\Big(\!\!\!\prod_{\footnotesize{\barray{c} m=0 \\ m \neq n \farray}}^{N-1}
{\tau - \tau_m \over
\tau_n - \tau_m}\Big) \qquad \mbox{for $\tau \in [0,U)$}~, \label{ai} \feq
with $\tau^i_n$ and $\aizn$ as in Eqs. \rref{taun} \rref{ain}.
By construction, the inequality \rref{faz}
$$ \left| \Big(s(\J(\tau),\te) - \R(\tau) s(I_0, 0) - \K(\tau) \Big)^i \right |
\leqs \aiz(\tau) \qquad \mbox{for all $\te \in \bT$} $$
is fulfilled, up to a small error, at all points $\tau = \tau_n$; we assume the same to happen everywhere
in $[0,U)$. Of course, being $C^{\infty}$, the functions \rref{ai} fulfill the regularity conditions
for application of the basic Proposition \ref{proprinc}.
\par
The above scheme (i)...(iv) has been employed to construct the functions $\aiz$ ($i=\p,\e,\y$) in the numerical
examples of Section \ref{seces} on the motions of satellites.
In these examples, $Q=30$ and $N=100$ (with such an $N$, the Lagrange interpolating
polynomials are computed very quickly by MATHEMATICA).
Admittedly, the approach followed does not account for the small errors mentioned in items (iii) (iv):
throughout the paper these errors, and the ones produced by the approximate inversion
of the matrix $1 - \ep \partial \alpha/\partial r$, are the only ones for which
an analytical estimate is not provided. \par
\section{Appendix. The functions $\boma{a^i_j}$,
$\boma{b^{i}}$,..., $\boma{e^{i}_{j k}}$ for the satellite problem.}
\label{appeae}
\salto
As usually, the indices $i,j,k$ range in $\{\p,\e,\y\}$. \salto
\textbf{Finding $\boma{a^{i}_{j}}$.}
As an example, let us illustrate the determination of $a^{\p}_{\y}$. To this purpose,
we use Eq. \rref{touses} with $i= \p, j = \y$, $I = \J(\tau) = (P_0, E_0, \J^\y(\tau))$ and
$\delta I = \delta J$, giving
\beq \SS^{\p}_{\y}(\J(\tau),\delta J, \te)  = \!
\int_{0}^1 \! \! \! d \var {\partial s^{\p}
\over \partial Y} (P_0 + \var \delta J^\p, E_0 + \var \delta J^\e, \J^\y(\tau)+ \var \delta J^\y, \te) \feq
$$ = \! \! \! \int_{0}^1 \! \! \! d \var {E_0 + \var \delta J^\e \over P_0 + \var \delta J^\p}~\Big(
3 \sin(\te + \J^\y(\tau) + \var \delta J^\y) - \sin(3 \te - \J^\y(\tau) - x \delta J^\y) \Big)~; $$
in the last passage, the derivative $\partial s^{\p}/\partial Y$ has been computed
from Eq. \rref{sx}.
Now we take the absolute value in the last equation, and use the elementary inequalities $|\int_{0}^1 d \var ...| \leqs
\int_{0}^1 d \var | ... |$,
$|E_0 + \var \delta J^\e| \leqs E_0 + | \delta J^\e |$,
$1/|P_0 + \var \delta J^\p| \leqs 1/(P_0 - |\delta J^\p|)$, $|\sin(..)| \leqs 1$. In this way we get
\beq | \SS^{\p}_{\y}(\J(\tau),\delta J, \te) | \leqs
4 {E_0 + |\delta J^\e| \over P_0 - | \delta J^\p|\, }
= a^{\p}_{\y}(r) \big|_{r^i = |\delta J^i|\, }~, \feq
where $a^{\p}_{\y}(r) := 4 (E_0 + \delta r^\e)/(P_0 - r^\p)$; this definition of
$a^{\p}_{\y}$ agrees with Eq. \rref{axx} (where we use the abbreviations $E_{+} := E_0 + r^\e$,
$P_{-} := P_0 - r^{\p}$). \salto
\textbf{Finding $\boma{b^i}$ and $\boma{c^i}$.}
We illustrate the determination of $b^\y$. To this purpose, we must first compute
the $\p$ component of the function on the left-hand side of Eq. \rref{fb} (for
$\J = (\J^i)$ and $\delta J = (\delta J^i)$ $(i=\p,\e,\y)$. Recalling again that $\J^\p(\tau) = P_0$
and $\J^\e(\tau) = E_0$ for all $\tau$, one finds
\beq \big( \w(\J(\tau) +\delta J, \te) - \FF(\J(\tau)) \, \v(\J(\tau) +
\delta J, \te) \big)^\y  \feq
$$ = {1 \over 512 {P_0}^3 (P_0 + \delta J^\p)^4 (E_0 + \delta J^\e)^{2} }~
\sum_{k=0,...,10; \ell =-6,...,6} P_{k \ell} \sin(k \vartheta + \ell (\J^\y(\tau) + \delta J^\y))~, $$
where the coefficients $P_{k \ell}$ are polynomials in $P_0, P_0 + \delta J^\p,
E_0 + \delta J^\e$; for example,
\beq P_{0 1} = 256 (E_0 + \delta J^\e) \big(-49 P_0^3 + 13 P_0^3 (E_0 + \delta J^\e)^2 -
16 (P_0 + \delta J^\p)^3 (E_0 + \delta J^\e)^2 \big)~. \feq
This implies
\beq \Big| \Big(\w(\J(\tau) +\delta J, \te) - \FF(\J(\tau)) \, \v(\J(\tau) +
\delta J, \te) \Big)^\y  \Big| \feq
$$ \leqs {1 \over 512 (P_0 - | \delta J^\p|)^3 (E_0 - | \delta J^\e|)^{2} }~
\sum_{k=0,...,10; \ell =-6,...,6} | P_{k \ell} |~; $$
the absolute value of each coefficient $P_{k \ell}$ is easily bounded, using
fairly rough estimates such as $|P_{0 1} | \leqs
256 (E_0 + |\delta J^\e|) \big( 49 P_0^3 + 13 P_0^3 (E_0 + |\delta J^\e|)^2 +
16 (P_0 + |\delta J^\p|)^3 (E_0 + |\delta J^\e|)^2 \big)$. The conclusion is
\beq \Big| \w(\J(\tau) +\delta J, \te) - \FF(\J(\tau)) \, \v(\J(\tau) +
\delta J, \te) \Big|^\y \leqs b^\y(r) \big|_{r^i = |\delta J^i|}~, \feq
where $b^\y$ is the function appearing in Eq. \rref{bx} (which is independent of $r^\y$). The determination
of $b^\p, b^\e$ and $c^\p, c^\e, c^\y$ is performed similarly, yielding Eqs. \rref{bx} \rref{cx}.
\salto
\textbf{Finding $\boma{d^{i}_{j}}$.}
As an example, let us consider the determination of $d^{\p}_{\y}$. We use
Eq. \rref{touse} with $i=\p, j = \y$, $I = \J(\tau) = (P_0, E_0, \J^\y(\tau))$ and
$\delta I = \delta J$, giving
\beq \GG^{\p}_{\y}(\J(\tau),\delta J)  = \!
\int_{0}^1 \! \! \! d \var {\partial \overline{p}^{\, \p}
\over \partial Y} (E_0 + \var \delta J^\e, P_0 + \var \delta J^\p,
\J^\y(\tau)+ \var \delta J^\y)  \feq
$$ = - 3 \pi\,
\int_{0}^1 \! \! \! d \var {(E_0 + \var \delta J^\e)^2
\over (P_0 + \var \delta J^\p)^3} \, \cos(2 \J^\y(\tau) + 2 \var \delta J^\y)~;
$$
in the last passage, the derivative $\partial \overline{p}^{\, \p}/\partial Y$ has been computed
from Eq. \rref{compp1}. The last equation implies
\beq | \GG^{\p}_{\y}(\J(\tau),\delta J) | \leqs 3 \pi \,
{(E_0 + |\delta J^\e|)^2 \over (P_0 - |\delta J^\p|)^3} = d^{\p}_{\y}(r) \big|_{r^i = |\delta J^i|}~, \feq
with $d^{\p}_{\y}$ as in Eq. \rref{dxx}.
\salto
\textbf{Finding $\boma{e^{i}_{j k}}$.} These functions
should bind the absolute values of the components $\HH^{i}_{j k}(\J(\tau), \delta J)$ of $\HH$;
these are provided by Eq. \rref{he1}, with $I = \J(\tau)$ and $\delta I = \delta J$.
From the cited equation, we see that we can take
$e^{i}_{j k}(\tau,r) := 0$ for $(i,j,k) \neq (\y,\p, \p)$. Again from \rref{he1}, we infer
\beq | \HH^{\y}_{\p \p}(\J(\tau), \delta J)| =
36 \pi \left| \int_{0}^1 d \var {(1 - \var) \over (P_0 + \var \delta J^\p)^4} \right| \feq
$$ \leqs 36 \pi  \, {\int_{0}^1 d \var (1 - \var) \over (P_0 - |\delta J^\p|)^4}
= {18 \pi \over (P_0 - |\delta J^\p|)^4} = e^{\y}_{\p \p}(r) \Big|_{r^i = |\delta J^i|}, $$
with $e^{\y}_{\p \p}$ as in Eq. \rref{eyh}~.
\salto
\section{Appendix. The approximate inverse \rref{invapp}.}
\label{appinv}
We must invert the $3 \times 3$ matrix $1 - \ep ({\partial \alpha/\partial r})(r)$, whose
elements are given by Eq. \rref{matel}. For $(\ep, r) \vain 0$, this equation implies
\beq 1 - \ep {\partial \alpha \over \partial r}(r) = 1 - \ep M - \ep r^k N_{(k)} - \ep^2 Z  +
O_3(\ep, r) \qquad \mbox{for $(\ep, r) \vain 0$}~,\label{ad1} \feq
where we provisionally write $M, N_{(k)}$ and $Z$ for the matrices of elements
\beq M^{i}_{j} := a^{i}_{j}(0)~, \quad
{N_{(k)}}^{i}_{j} := {\partial a^{i}_{k} \over \partial r^j}(0) + {\partial a^{i}_{j} \over \partial r^k}(0)~,
\quad Z^{i}_{j} := {\partial b^{i} \over \partial r^j}(0)~, \quad \label{add1} \feq
($i, j, k \in \{\p,\e,\y\}$). Eq. \rref{ad1}, with the standard $X \vain 0$
expansion $(1 - X)^{-1} = 1 + X + X^2 + O_3(X)$, implies
$(1 - \ep ({\partial \alpha/\partial r})(r))^{-1} = $
$1 + (\ep M  + \ep r^k N_{(k)} + \ep^2 Z )$ +$ (\ep M + \ep r^k N_{(k)} + \ep^2 Z )^2 +
O_3(\ep, r) = $ $1 + \ep M + \ep r^k N_{(k)} + \ep^2 Z  + \ep^2 M^2 + O_{3}(\ep, r)$. Summing up,
\beq \Big(1 - \ep {\partial \alpha \over \partial r}(r)\Big)^{-1} \!\!=
1 + \ep M  + \ep r^k N_{(k)} + \ep^2 \Q + O_3(\ep, r)~, \qquad \Q
:= Z + M^2~. \label{ad2} \feq This justifies the approximate
inverse \rref{invapp} if we check that the matrices $M, N_{(k)}$
and $\Q$, defined here via Eq.s \rref{add1} \rref{ad2}, coincide
with the homonymous matrices in Eq.s \rref{emme} \rref{enne}
\rref{qu}. This is obtained substituting in \rref{add1} \rref{ad2}
the explicit expressions \rref{axx} \rref{bx} of the functions
$a^i_{j}$ and $b^i$. \vfill \eject \noindent
\textbf{Acknowledgments.} This work has been partially supported
by the GNFM of Istituto Nazionale di Alta Ma\-te\-ma\-ti\-ca and
by MIUR, Research Project Cofin/2006 "Metodi geometrici nella
teoria delle onde non lineari e applicazioni". \salto

\end{document}